\documentclass[draftclsnofoot,onecolumn,12pt]{IEEEtran}  

\usepackage[font=small,labelfont=bf]{caption}
\usepackage{subfig}
\usepackage{color}
\usepackage[final]{graphicx}
\usepackage{bm}
\usepackage{cite}
\usepackage{amssymb,amsmath,amsthm}
\usepackage[normalem]{ulem}
\usepackage{paralist}
\usepackage[acronym]{glossaries}
\usepackage[bookmarks]{hyperref}
\usepackage{tikz}
\usetikzlibrary{shapes,arrows}

\newacronym{sat}{SAT}{Satellite}
\newacronym{rhs}{r.h.s.}{right hand side}
\newacronym{sud}{SUD}{single-user decoding}
\newacronym{iid}{i.i.d.}{identical independent distributed}
\newacronym{jt}{JT}{joint transmission}
\newacronym{mimo}{MIMO}{Multiple input multiple output}
\newacronym{csi}{CSI}{channel-state information}
\newacronym{csit}{CSIT}{channel-state information at the transmitter}
\newacronym{csir}{CSIR}{channel-state information at the receiver}
\newacronym{leo}{LEO}{low-earth-orbit}
\newacronym{snr}{SNR}{signal-to-noise  ratio}
\newacronym{awgn}{AWGN}{additive white complex Gaussian noise}
\newacronym{los}{LOS}{line-of-sight}
\newacronym{pll}{PLL}{Phase-Lock-Loop}
\newacronym{dpll}{DPLL}{discrete-phase-lock-loop}
\newacronym{lt}{LT}{land terminal}
\newacronym{lpf}{LPF}{low-pass-filter}
\newacronym{zf}{ZF}{zero-forcing}
\newacronym{vco}{VCO}{voltage-controlled-oscillator}
\newacronym{uca}{UCA}{uniform circular array}
\newacronym{cdf}{CDF}{cumulative distribution function}
\newacronym{mse}{MSE}{mean squared error}
\newacronym{mnsll}{MSL}{mean number of samples to lose lock}
\newacronym{tdd}{TDD}{time division duplex}
\newacronym{fdd}{FDD}{frequency division duplex}
\newacronym{nc}{NC}{non-cooperative}

\newacronym{sc}{SatCom}{Satellite communication}
\newacronym{mu}{MU-\gls{mimo}}{multi-user \gls{mimo}}

\newtheorem{theorem}{Theorem}

\newtheorem{lemma}[theorem]{Lemma}

\newtheorem{claim}[theorem]{Claim}

\newtheorem{definition}{Definition}

\newcommand{\beq}{\begin{equation}}
\newcommand{\eeq}{\end{equation}}
\newcommand{\bea}{\begin{array}}
\newcommand{\ena}{\end{array}}
\newcommand{\bds}{\begin {itemize}}
\newcommand{\eds}{\end {itemize}}
\newcommand{\bdf}{\begin{definition}}
\newcommand{\blm}{\begin{lemma}}
\newcommand{\edf}{\end{definition}}
\newcommand{\elm}{\end{lemma}}
\newcommand{\bthm}{\begin{theorem}}
\newcommand{\ethm}{\end{theorem}}
\newcommand{\bprp}{\begin{prop}}
\newcommand{\eprp}{\end{prop}}
\newcommand{\bcl}{\begin{claim}}
\newcommand{\ecl}{\end{claim}}
\newcommand{\bcr}{\begin{coro}}
\newcommand{\ecr}{\end{coro}}
\newcommand{\bquest}{\begin{question}}
\newcommand{\equest}{\end{question}}

\newcommand{\vct}[1]{\mbox{\boldmath$ \bf #1$}}

\newcommand{\beqna}{\begin{eqnarray}}
\newcommand{\eeqna}{\end{eqnarray}}

\begin{document}

\title{
\textbf{Pseudo channel reciprocity in FDD satellite channels}\\
}


\author{Rei Richter, \textit{Student, IEEE}, Itsik Bergel, \textit{Senior, IEEE}, Yair Noam, \textit{Member, IEEE}, Ephi Zehavi, \textit{Fellow, IEEE}}

\maketitle

\begin{abstract}
Channel reciprocity can significantly reduce the overhead of obtaining channel-state information at the transmitter (CSIT). However, true reciprocity only exists in time division duplex (TDD). In this paper, we propose a novel tracking method that exploits implicit reciprocity in FDD line-of-sight (LOS) channels as in low-earth-orbit (LEO) satellite communication (SatCom). This channel reciprocity, dubbed pseudo-reciprocity, is crucial for applying multiple-user multiple-input multiple-output (MU-MIMO) to SatCom, which requires CSIT. We consider an LEO SatCom system where multiple satellites communicate with a multi-antenna land terminal (LT). In this innovative method, the LT can track the downlink channel changes and use them to estimate the uplink channels. The proposed method achieves precoding performance that is comparable to precoding with full CSIT knowledge. Furthermore, the use of pseudo reciprocity typically requires only initial CSIT feedback when the satellite rises. Over very long periods of time, pseudo reciprocity can fail as a result of phase ambiguity, which is referred to as cycle slip. We thus also present a closed-form approximation for the expected time until cycle slip, which indicates that in normal operating conditions, these cycle slips are extremely rare. Our numerical results provide strong support for the derived theory.
\end{abstract}

\section{Introduction}

\gls{sc} systems are essential for universal coverage since they do not require  terrestrial infrastructure. For this reason, commercial and public institutions plan to deploy thousands of \gls{leo} satellites over the next decade. These systems must have  high throughput to justify such a massive investment. \gls{mimo} technology and in particular \gls{mu} are vital for overcoming this challenge in \gls{sc} \cite{richter2020downlink, goto2018leo, gao2021sum}. \par
\gls{mu} can dramatically enhance throughput, but requires both \gls{csir} and \gls{csit} \cite{richter2015optimal, arad2022c}. The receiver usually estimates the \gls{csir} through received pilot sequences. Obtaining \gls{csit}, on the other hand, is more challenging.
\par
In \gls{tdd} systems, CSIT can be obtained using  channel reciprocity (e.g., \cite{liu2015performance, jiang2015mimo}). In these systems,  the uplink and downlink channels are in the same frequency band and hence experience the same channel (reciprocal). Thus, the channel estimated during the receive time slot (as \gls{csir}) can be used in the transmit time slot (as \gls{csit}). Note that the receive hardware and transmit hardware are not reciprocal. Thus, the use of channel reciprocity requires hardware  calibration \cite{jiang2018framework,rogalin2014scalable}.
\par
In \gls{fdd} systems,  the uplink and downlink channels use different  frequencies and hence channel reciprocity does not hold. Note that the uplink and downlink channels typically have some common properties, such as angular power spectrum, spatial covariance matrix, delay, and multipath angles \cite{zhong2020fdd,yin2022partial}.  These reciprocal properties can be used for user selection \cite{zhang2021scalable} or to reduce the feedback rate \cite{kim2019path,lin2021deep,liu2021hyperrnn}. However, since the channel gains are not reciprocal, all these schemes still require significant feedback.
\par
Although channel reciprocity in FDD systems may initially appear unattainable, it is, in fact, possible within SatCom systems. The SatCom channel is predominantly characterized by a LOS component, which changes over time. This LOS channel allows for the inference of one link direction based on the other, thereby enabling channel estimation through reciprocity.  Consequently,  SatCom systems offer a unique opportunity for exploiting channel reciprocity in the FDD setup.\par
In a LOS channel, the channel phase is determined by the delay and the carrier frequency\footnote{The effects of transmitter hardware, antenna gains, and atmospheric and rain attenuation obviously affect the phase. But, we are interested in the phase changes between the antennas in the same equipment, which experience the same effects.}. Thus, exact  knowledge of the delay is sufficient to achieve CSI. But in most practical scenarios, it is not feasible to estimate the delay with sufficient accuracy. Instead, the channel phase can be accurately estimated from the CSIR up to a modulo $2\pi$. In TDD systems, where   the carrier frequency is similar in both link directions, the  channel phases are also identical. Hence, a good estimation of the modulo $2\pi$ phase is sufficient to be able to tap channel reciprocity.
\par
However, achieving channel reciprocity in FDD systems is more complicated and not as straightforward as in TDD systems. This is primarily because the uplink and downlink channels have  different frequencies, resulting in unequal channel phases.  Thus, knowledge of the phase modulo $2\pi$ in one link direction is not enough for phase estimation in the other direction. So far, there is no known technique for exploiting channel reciprocity in FDD \gls{leo} \gls{sc}. 
\par
This paper proposes an innovative scheme for exploiting channel reciprocity in \gls{leo} \gls{sc} systems which typically have dominant \gls{los} and employ \gls{fdd}. In the proposed scheme, the \gls{lt} estimates its uplink \gls{csit} from its downlink \gls{csir}, and requires only \gls{csi} feedback  for initialization. Our main result is based on the fact that the Doppler shift in the uplink and downlink is identical up to a multiplicative constant. Thus, using appropriate techniques, the phase variations of the uplink channel can be predicted from the estimated phase variations in the downlink channel.

We consider a network containing several \gls{leo} satellites communicating with a single multi-antenna \gls{lt}. The \gls{lt} estimates the downlink channels (\gls{csir}) via pilot sequences sent by the satellites. From this \gls{csir}, the \gls{lt} calculates its received beamformer for extracting each satellite's data stream. Moreover, the \gls{lt} uses its \gls{csir} to track the uplink channel (\gls{csit}) by tracking the downlink channel phase variations. Finally, the \gls{lt} uses this \gls{csit} to calculate a \gls{zf} precoding matrix for sending data simultaneously to the satellites in the uplink.
\par
\textit{Notations:} We denote Matrices and vectors
by boldface upper and lower case letters, respectively. We use  $()^*$, and $()^H$ for  the conjugate and
transpose operations, respectively, and
 $\mathbb{E}(\cdot)$
and $\rm{Var}(\cdot)$, for the expectation and variance operations, respectively.  Let $y\in{\mathbb R}$, then we defined its $2\pi$ modulo as
\begin{IEEEeqnarray}{rCl}
[y]_{2\pi}=((y+\pi)\mod 2\pi)-\pi, \qquad-\pi\leq [y]_{2\pi}\leq\pi.
\end{IEEEeqnarray}
For $x\in{\mathbb C}$, we denote its phase modulo \( 2\pi\) by $\measuredangle x$.

\section{System Model}\label{sec:system_model}
We consider a \gls{fdd} satellite communication network where a fixed land terminal (LT) with $M$
antennas communicates simultaneously with $L\le M$  \gls{leo} satellites, each having a single antenna. In the downlink, all satellites send pilots periodically whereas in the uplink, whenever the LT transmits, it sends independent data streams simultaneously to the satellites.
\begin{figure}[t]
\begin{center}
\includegraphics[width=90mm]{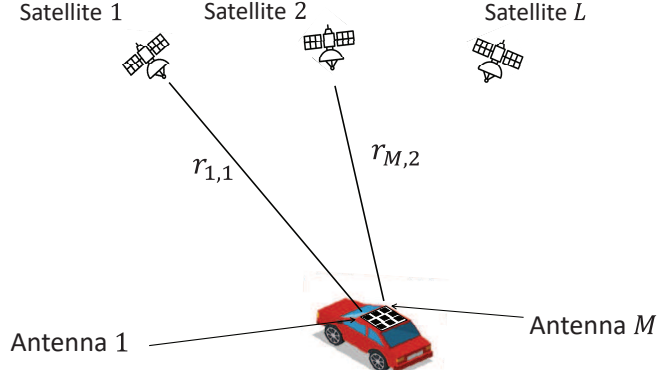}
    \caption{Illustration of the  system model}
\label{fig_1}
\end{center}
\end{figure}
The LT tracks the downlink channel phase variations from the pilot sent by each satellite. Based on this \gls{csir}, the LT extracts each satellite data stream using received beamforming.
In the uplink, on the other hand, the LT uses transmit-beamforming to null interference from signals designated to other satellites.  To do so, the LT needs to extract the uplink \gls{csit} from the downlink \gls{csir}.

\subsection{Downlink}\label{sec:sync}

The LT uses the pilot symbols transmitted by the satellites to construct a matrix channel estimation. The channel estimation for the $n$-th symbol can be written as:

 \begin{IEEEeqnarray}{rCl}\label{eq:tilde_HD1}
\hspace{-10mm}\mathbf{\tilde H}^{\mathrm{D}}[n]=\begin{bmatrix}\frac{\Upsilon^\mathrm{D}_{1}}{r_{1,1}[n]}e^{-j2\pi (f^\mathrm{D}_1 +f_{\mathrm{c}}^\mathrm{D})r_{1,1}[n]/c}e^{j2\pi f^\mathrm{D}_{1} nT}e^{j\alpha^\mathrm{D}_1}  & \cdots & \frac{\Upsilon^\mathrm{D}_{1}}{r_{1,L}[n]}e^{-j2\pi (f^\mathrm{D}_L +f_{\mathrm{c}}^\mathrm{D})r_{1,L}[n]/c}e^{j2\pi f^\mathrm{D}_{L} nT}e^{j\alpha^\mathrm{D}_L} \\
\vdots\ & \ddots & \vdots \\
\frac{\Upsilon^\mathrm{D}_{1}}{r_{1,1}[n]}e^{-j2\pi (f^\mathrm{D}_1 +f_{\mathrm{c}}^\mathrm{D})r_{M,1}[n]/c}e^{j2\pi f^\mathrm{D}_{1} nT}e^{j\alpha^\mathrm{D}_1} & \cdots & \frac{\Upsilon^\mathrm{D}_{1}}{r_{1,L}[n]}e^{-j2\pi (f^\mathrm{D}_L +f_{\mathrm{c}}^\mathrm{D})r_{M,L}[n]/c}e^{j2\pi f^\mathrm{D}_{L} nT}e^{j\alpha^\mathrm{D}_L}  \\
\end{bmatrix}+\mathbf{N}^{\mathrm{D}}[n]\hspace{7mm}
\end{IEEEeqnarray}
where $f^\mathrm{D}_c$ is the LT downlink carrier frequency, $f^\mathrm{D}_\ell$ is the satellite downlink hardware frequency-offset,
$r_{m,\ell}[n]$ is the distance between satellite-$\ell$ and the $m$-th LT antenna  at sample $n$, $\alpha^\mathrm{D}_\ell$ is the phase offset related to satellite $\ell$, $T$ is the  symbol duration, $c$ is the speed of light and $\Upsilon^\mathrm{D}_{\ell}$ is a constant expressing the effect of transmitter hardware, antenna gains, and atmospheric and rain attenuation
for satellite $\ell$ \footnote{The channel factoring in \eqref{eq:tilde_HD1} includes the standard approximation, where we assume that $(r_{m,\ell}[n]-r_{k,\ell}[n])$ is negligible compared to $r_{m,\ell}[n]$ but $(r_{m,\ell}[n]-r_{k,\ell}[n])/ f^{\mathrm{D}}_c$ is not negligible.}. The estimation error term, $\mathbf{N}^{\mathrm{D}}[n]\in\mathbb{C}^{M\times L}$ is composed of \gls{iid} complex Gaussian random variables with zero mean and variance of $\sigma^2$.

The matrix $\mathbf{\tilde H}^\mathrm{D}[n]$ also can be written as
 \begin{IEEEeqnarray}{rCl}\label{eq:tilde_HD}
\mathbf{\tilde H}^{\mathrm{D}}[n]&=&\mathbf{H}^\mathrm{D}[n]\mathbf{D}^\mathrm{D}[n]+\mathbf{N}^{\mathrm{D}}[n]
\end{IEEEeqnarray}
where
\begin{IEEEeqnarray}{rCl}\label{eq:H_D}
\mathbf{H}^\mathrm{D}[n]=\begin{bmatrix}e^{j\Theta^\mathrm{D}_{1,1}[n]}  & \cdots & e^{j\Theta^\mathrm{D}_{1,L}[n]} \\
\vdots\ & \ddots & \vdots \\
e^{j\Theta^\mathrm{D}_{M,1}[n]} & \cdots & e^{j\Theta^\mathrm{D}_{M,L}[n]}  \\
\end{bmatrix}
\end{IEEEeqnarray}

contains only phase information, where
\begin{IEEEeqnarray}{rCl}\label{eq:DL_true_phase}
\Theta^\mathrm{D}_{m,\ell}[n]&=&-\frac{2\pi (f^\mathrm{D}_\ell +f_{\mathrm{c}}^\mathrm{D})r_{m,\ell}[n]}{c}+2\pi f^\mathrm{D}_{\ell} nT+\alpha^\mathrm{D}_\ell
\end{IEEEeqnarray}
and
\begin{IEEEeqnarray}{rCl}\label{eq:D_D}
\hspace{-1mm}\mathbf{D}^\mathrm{D}[n]=\begin{bmatrix}\frac{\Upsilon^\mathrm{D}_{1}}{r_{1,1}[n]}   & \cdots & 0 \\
\vdots\ & \ddots & \vdots \\
0 & \cdots & \frac{\Upsilon^\mathrm{D}_{L}}{r_{1,L}[n]}   \\
\end{bmatrix}
\end{IEEEeqnarray}
 is a diagonal matrix containing the amplitude for  all channel entries corresponding to the same satellite.
\par
Note that matrix $\mathbf{D}^\mathrm{D}[n]$ changes very slowly, and much slower than matrix $\mathbf{H}^\mathrm{D}[n]$.   Therefore,  the LT has a long time to estimate the matrix $\mathbf{D}^\mathrm{D}[n]$, and we assume that this estimation results in a negligible estimation error.

\subsection{Uplink}
The LT transmits $L$ independent data streams, each designated for a distinct satellite. The baseband output signal at antenna $m\in\{1,...,M\}$ is
\begin{IEEEeqnarray}{rCl}
        g_m(t) &=& \sum_n x_{m}[n] p(t-nT)
\end{IEEEeqnarray}
where $x_m[n]$ is the $n$-th symbol transmitted by antenna $m$  and $p(t)$ is the normalized pulse shape, which is free of inter symbol interference (ISI). Thus, the pulse auto-correlation satisfies  $R_p(n T)= \int p(t) p^*(t-nT) dt=\delta[n]$. The corresponding passband signal is
\begin{IEEEeqnarray}{rCl}
        g^{\mathrm{PB}}_m(t) &=& \Re\left\{\sqrt{2}\sum_n x_{m}[n] p(t-nT_{\ell})e^{j2\pi f_c^\mathrm{U} t}\right\}\label{passband_signal}
\end{IEEEeqnarray}
where $f^\mathrm{U}_{\text c}$ is the uplink carrier frequency. The  received passband signal  at satellite $\ell$ is
\begin{IEEEeqnarray}{rCl}
        \label{EqRecivedSignal}
        y^{\mathrm{PB}}_{\ell}(t) &=&\sum_{m=1}^M\frac{\Upsilon^\mathrm{U}_{\ell}}{r_{m,\ell}(t)}  g^{\mathrm{PB}}_m\left(t- \frac{r_{m,\ell}(t)}{c}\right)e^{j\alpha^\mathrm{U}_\ell} +n_\ell(t)
\end{IEEEeqnarray}
where  $n_\ell(t)$ is a complex circular white Gaussian noise with two sided spectral density  $\sigma^2_{\mathrm{U}}$, and
 $\Upsilon^\mathrm{U}_{\ell}$ is a constant expressing the effect of transmitter hardware, antenna gains, and atmospheric and rain attenuation. The received baseband signal at satellite $\ell$ after multiplying by $\sqrt{2}e^{-j2\pi (f_c^\mathrm{U}+f_\ell^\mathrm{U})t}$ and applying LPF  is
\begin{IEEEeqnarray}{rCl}
        \label{EqRecivedSignal}
        y^\mathrm{U}_{\ell}(t) 
&=&\sum_{m=1}^M\frac{\Upsilon^\mathrm{U}_{\ell}e^{j\alpha^\mathrm{U}_\ell}}{r_{m,\ell}(t)}  \sum_n x_{m}[n] p(t-nT_{\ell}- \frac{r_{m,\ell}(t)}{c})e^{-j2\pi f_c^\mathrm{U}  \frac{r_{m,\ell}(t)}{c}} e^{-j2\pi f_\ell^\mathrm{U}t}+n_\ell(t).\notag
\end{IEEEeqnarray}


\par
Using the standard approximation again, we can replace $r_{m,\ell}(t)$ by $ r_{1,\ell}(t)$ in the argument of $p(t)$ (but not in the exponent). In addition, the sampling instances at the output of the match filter on satellite $\ell$ are $nT+\tau_\ell, n\in{\mathbb Z}$ where $\tau_\ell=\frac{r_{1,\ell}(nT)}{c}$. The resulting samples are
\begin{IEEEeqnarray}{rCl}
y^\mathrm{U}_{\ell}[n] &=& y^\mathrm{U}_{\ell}(t)*{p^*(-t)}\Big|_{t=nT+\tau_\ell}\notag\\
&=&\int_{-\infty}^\infty \sum_{m=1}^M\frac{\Upsilon^\mathrm{U}_{\ell}e^{j\alpha^\mathrm{U}_\ell}}{r_{m,\ell}(nT)} \sum_{n'} x_{m}[n'] p(\tau-n'T_{\ell}- \tau_\ell)e^{-j2\pi f_c^\mathrm{U}  \frac{r_{m,\ell}(nT)}{c}} e^{-j2\pi f_\ell^\mathrm{U}t}p^*(\tau-nT-\tau_\ell)d\tau+n_{\ell}^\mathrm{U}[n].\notag
\end{IEEEeqnarray}
As the pulse shape $p(t)$ is normalized and ISI free, the noise $n^\mathrm{U}_{\ell}[n]$ is a complex circular white Gaussian with variance $\sigma^2_{\mathrm{U}}$.

We assume that the gap between the distances from the $\ell$ satellite to the LT first antenna and to its $m$-th antenna, $|r_{m,\ell}(nT) - r_{1,\ell}(nT)|$, is neglected in the attenuation term $(\frac{1}{r_{1,\ell}(nT)}\approx\frac{1}{r_{m,\ell}(nT)})$. This small gap only affects the argument of the exponent, where it is multiplied by $f^\mathrm{U}_c r_{m,\ell}(nT)/c$, hence
\begin{IEEEeqnarray}{rCl}
y^\mathrm{U}_{\ell}[n]
&\approx& \sum_{m=1}^M \frac{\Upsilon^\mathrm{U}_{\ell}e^{j\alpha^\mathrm{U}_\ell}}{r_{1,\ell}(nT)} x_{m}[n]e^{-j2\pi f_c^\mathrm{U}  \frac{r_{m,\ell}(nT)}{c}} e^{-j2\pi f_\ell^\mathrm{U}nT}+n_{m,\ell}^\mathrm{U}[n].
\end{IEEEeqnarray}

\par
Next, we denote $\mathbf{n}^\mathrm{U}[n]=\left[n^\mathrm{U}_{0}[n],\ldots,n^\mathrm{U}_{L-1}[n]\right]^T$, and $\mathbf{x}^\mathrm{U}_\ell[n]=\left[x^\mathrm{U}_{0}[n],\ldots,x^\mathrm{U}_{M-1}[n]\right]^T$
and $\mathbf{y}^\mathrm{U}[n]=\left[y^\mathrm{U}_{1}[n],\ldots,y^\mathrm{U}_{L}[n]\right]^T$.
Thus, $\mathbf{y}^\mathrm{U}[n]$  can be written as
\begin{IEEEeqnarray}{rCl}\label{eq:uplink_model}
\mathbf{y}^\mathrm{U}[n]&=&(\tilde{\mathbf{H}}^\mathrm{U}[n])^\mathrm{T} \mathbf{x}^\mathrm{U}_\ell[n]+\mathbf{n}^\mathrm{U}[n] \end{IEEEeqnarray}
where
\begin{IEEEeqnarray}{rCl}
\hspace{-1mm}\tilde{\mathbf{H}}^\mathrm{U}[n]=\begin{bmatrix}\frac{\Upsilon^\mathrm{U}_{1}}{r_{1,1}(t)}  e^{-j2\pi f_c^\mathrm{U}  \frac{r_{1,1}(nT)}{c}} e^{-j2\pi f_1^\mathrm{U}nT}e^{j\alpha^\mathrm{U}_1}  & \cdots &  \frac{\Upsilon^\mathrm{U}_{L}}{r_{1,L}(t)}  e^{-j2\pi f_c^\mathrm{U}  \frac{r_{1,L}(nT)}{c}} e^{-j2\pi f_L^\mathrm{U}nT}e^{j\alpha^\mathrm{U}_L}\\
\vdots\ & \ddots & \vdots \\
\frac{\Upsilon^\mathrm{U}_{1}}{r_{1,1}(t)}  e^{-j2\pi f_c^\mathrm{U}  \frac{r_{M,1}(nT)}{c}} e^{-j2\pi f_1^\mathrm{U}nT}e^{j\alpha^\mathrm{U}_1} & \cdots & \frac{\Upsilon^\mathrm{U}_{L}}{r_{1,L}(t)}  e^{-j2\pi f_c^\mathrm{U}  \frac{r_{M,L}(nT)}{c}} e^{-j2\pi f_L^\mathrm{U}nT}e^{j\alpha^\mathrm{U}_L}  \\
\end{bmatrix}\notag.
\end{IEEEeqnarray}
Note that we write \eqref{eq:uplink_model} with a transpose on the channel matrix for convenience (which is common in reciprocity models).
\par
Matrix $\mathbf{\tilde H}^\mathrm{U}[n]$ can be written as \begin{IEEEeqnarray}{rCl}\label{eq:HU}
\mathbf{\tilde H}^{\mathrm{U}}[n]&=&\mathbf{H}^{\mathrm{U}}[n]\mathbf{D}^{\mathrm{U}}[n]
\end{IEEEeqnarray}
where
\begin{IEEEeqnarray}{rCl}
\hspace{-1mm}\mathbf{D}^\mathrm{U}[n]=\begin{bmatrix}\frac{\Upsilon^\mathrm{U}_{1}}{r_{1,1}(t)}e^{-j2\pi f_1^\mathrm{U}nT}e^{j\alpha^\mathrm{U}_1}  & \cdots & 0 \\
\vdots\ & \ddots & \vdots \\
0 & \cdots & \frac{\Upsilon^\mathrm{U}_{L}}{r_{1,L}(t)}e^{-j2\pi f_L^\mathrm{U}nT}e^{j\alpha^\mathrm{U}_L}  \\
\end{bmatrix}\notag.
\end{IEEEeqnarray}
 is a diagonal matrix containing the terms that are equal for all channel elements corresponding to the same satellite, and
\begin{IEEEeqnarray}{rCl}\label{eq:H_UL}
\mathbf{H}^\mathrm{U}[n]=\begin{bmatrix}e^{j\Theta^\mathrm{U}_{1,1}[n]}  & \cdots & e^{j\Theta^\mathrm{U}_{1,L}[n]} \\
\vdots\ & \ddots & \vdots \\
e^{j\Theta^\mathrm{U}_{M,1}[n]} & \cdots & e^{j\Theta^\mathrm{U}_{M,L}[n]}  \\
\end{bmatrix}
\end{IEEEeqnarray}
is a phase matrix, which takes into account the phase differences between LT antennas where
\begin{IEEEeqnarray}{rCl}\label{eq:UL_true_phase}
\Theta^\mathrm{U}_{m,\ell}[n]&=&-\frac{2\pi f_{\mathrm{c}}^\mathrm{U}r_{m,\ell}[n]}{c}.
\end{IEEEeqnarray}

\subsection{Uplink ZF precoding}

  In  the uplink, the LT employs a ZF precoding to cancel interferences between streams designated for different satellites. Let $s_\ell^\mathrm{U}[n]$ be the data symbol transmitted to satellite $\ell$ and
 $\vct{s}^\mathrm{U}[n] =[s_1^\mathrm{U}[n], . . . , s_L^\mathrm{U}[n]]^T$ be the vector of all transmitted symbols at time $n$. This vector is precoded using the precoding matrix, $\vct{T}[n]$, to generate:
$\vct{x}^\mathrm{U}[n] = \vct{T}[n]\vct{s}^\mathrm{U}[n]$
. The  ZF precoding matrix is then
\begin{IEEEeqnarray}{rCl}
\vct{T}[n] = \frac{\vct{H}^{\mathrm{U}}[n]^*((\vct{H}^\mathrm{U}[n])^T\vct{H}^{\mathrm{U}}[n]^*)^{-1}}{\sqrt{\mbox{Tr}[((\vct{H}^\mathrm{U}[n])^T\vct{H}^{\mathrm{U}}[n]^*)^{-1}]}}.
\label{eq:T_UL}
\end{IEEEeqnarray}
Note that the ZF precoding does not require full CSIT, $\tilde{\mathbf{H}}^\mathrm{U}[n]$, but only the phase matrix $\mathbf{H}^\mathrm{U}[n]$.
Using the ZF precoding,  the received sampled signals at the satellites are
\begin{IEEEeqnarray}{rCl}\label{eq:y_UL_precoding}
 \vct{y}^\mathrm{U}[n]&=& \frac{\vct{D}^\mathrm{U}[n]}{\sqrt{\mbox{Tr}[((\vct{H}^\mathrm{U}[n])^T\vct{H}^{\mathrm{U}}[n]^*)^{-1}]}}\vct{s}^\mathrm{U}[n]+ \vct{n}^\mathrm{U}[n].
 \end{IEEEeqnarray}

To summarize, the LT can cancel interference in downlink and uplink channels as long as it has a good enough estimate of $\vct{H}^\mathrm{D}[n]$ and $\vct{H}^\mathrm{U}[n]$.

\section{Pseudo channel reciprocity}\label{sec:pseudo}
The utilization of ZF, both in the downlink and  the uplink, requires CSI.  In the downlink, we assume that the LT  achieves \gls{csir} via pilot sequences. For the uplink ZF precoding, however, the LT needs CSIT, which is far more challenging than CSIR.  Without channel reciprocity, \gls{csit} requires channel estimation at the receiver and feedback, which incurs an overhead. Channel reciprocity is often used in \gls{tdd} systems, where both uplink and downlink use the same carrier frequency. In our case, which includes FDD, LOS and high Doppler shifts, achieving channel reciprocity is difficult. We now propose a new scheme that utilizes downlink channel estimation to track the uplink channel. While the scheme still includes uplink channel estimation at the satellites and feedback, it can be at a significantly lower rate.
\subsection{problem formulation}
As the matrices $\mathbf{D}^\mathrm{U}[n]$ and $\mathbf{D}^\mathrm{D}[n]$ are diagonal, both the ZF precoder and the ZF equalizer only require the knowledge of the matrices $\mathbf{H}^\mathrm{U}[n]$ and $\mathbf{H}^\mathrm{D}[n]$, respectively. Moreover, since all the elements in the matrices $\mathbf{H}^\mathrm{U}[n]$ and $\mathbf{H}^\mathrm{D}[n]$ have an amplitude of $1$, it is sufficient to study their phases. Thus, in the following, we focus on the channel phases; i.e., the phases of the elements of the matrices $\mathbf{H}^\mathrm{U}[n]$ and $\mathbf{H}^\mathrm{D}[n]$. Let \(h^\mathrm{U}_{m,\ell}[n]\) and \(h^\mathrm{D}_{m,\ell}[n]\) be the \( (m,\ell)\) entries of \(\mathbf{H}^\mathrm{U}[n]\) and \(\mathbf{H}^\mathrm{D}[n]\), respectively: recall that $\Theta^\mathrm{U}_{m,\ell}[n]$ and $\Theta^\mathrm{D}_{m,\ell}[n]$ are the corresponding phases.
\par
The channel estimation goal is to track $\Theta^\mathrm{U}_{m,\ell}[n]$ using sequential estimates of
$ \Theta^\mathrm{D}_{m,\ell}[n]$. The challenge of this
 reciprocity procedure  is  that the  LT does not observe estimates of the actual channel phase, but only its modulo $2\pi$.  Thus, the goal is actual work is to track $[\Theta^\mathrm{U}_{m,\ell}[n]]_{2\pi}$ given sequential estimates of $[\Theta^\mathrm{D}_{m,\ell}[n]]_{2\pi}$.

\subsection{Pseudo Reciprocity}
 We now propose a tracking algorithm wherein the LT  extracts  the uplink channel phases from the estimation of the downlink channel phases.
 For simplicity of presentation, in this section we present a scheme that only works when the   frequency offsets  between the satellites and the LT are negligible (i.e., we assume $f_\ell^\mathrm{D}=0$).  In the next section, we present  an algorithm that also works in the presence of such frequency offset.
 \par
We first present the scheme in the absence of estimation errors, but later present a performance analysis in the presence of noise. For $f_\ell^\mathrm{D}=0$, the measured channel phase is
\begin{IEEEeqnarray}{rCl}\label{eq:downlink_phase_fl_zero}
\measuredangle h^\mathrm{D}_{m,\ell}[n]&=&[\Theta^\mathrm{D}_{m,\ell}[n]]_{2\pi}=\left[-2\pi f_{\mathrm{c}}^\mathrm{D}r_{m,\ell}(nT)/c+\alpha_\ell^\mathrm{D}\right]_{2\pi}.
\end{IEEEeqnarray}
The LT has no information on $\alpha_\ell^D$.  Thus all our schemes will require the initial feedback of a single estimate of the uplink $ \Theta^\mathrm{U}_{m,\ell}[0]$. Once we have this feedback, we will show that it is sufficient to track \( \measuredangle h^\mathrm{D}_{m,\ell}[n]\)  to estimate  \(\measuredangle  h_{m,\ell}^{\rm U} [n]\).
\par
If the LT could obtain a reasonable estimate of ${2\pi f_{\mathrm{c}}^\mathrm{D}r_{m,\ell}(nT)}/{c}$,  it would extract $r_{m,\ell}(nT)$ to get a good estimate of $\measuredangle h^\mathrm{U}_{m,\ell}[n]$. However, the LT does not estimate ${2\pi f_{\mathrm{c}}^\mathrm{D}r_{m,\ell}(nT)}/{c}$ but rather its $2\pi$ modulo. To address this problem, we write \( \measuredangle h^\mathrm{D}_{m,\ell}[n]\) and  \(\measuredangle  h_{m,\ell}^{\rm U} [n]\) as sums over phase increments
\begin{IEEEeqnarray}{rCl}
\measuredangle h^\mathrm{D}_{m,\ell}[n]&=&\left[\Theta^\mathrm{D}_{m,\ell}[0]-\frac{2\pi f_{\mathrm{c}}^\mathrm{D}}{c}\sum_{k=1}^nr_{m,\ell}(kT)-r_{m,\ell}((k-1)T)\right]_{2\pi}\label{eq:explicit_DL}\\
\measuredangle h^\mathrm{U}_{m,\ell}[n]&=&\left[\Theta^\mathrm{U}_{m,\ell}[0]-\frac{2\pi f_{\mathrm{c}}^\mathrm{U}}{c}\sum_{k=1}^nr_{m,\ell}(kT)-r_{m,\ell}((k-1)T)\right]_{2\pi}\label{eq:explicit_UL}.
\end{IEEEeqnarray}
We define the downlink measured phase change as
\begin{IEEEeqnarray}{rCl}
\Delta^\mathrm{D}_{m,\ell}[n]&=&\left[\measuredangle h^\mathrm{D}_{m,\ell}[n]-\measuredangle h^\mathrm{D}_{m,\ell}[n-1]\right]_{2\pi}\notag\\\label{eq:downlink_delta_orig}
&=&-\left[\left[2\pi f_{\mathrm{c}}^\mathrm{D}r_{m,\ell}(nT)/c\right]_{2\pi}-\left[2\pi f_{\mathrm{c}}^\mathrm{D}r_{m,\ell}((n-1)T)/c\right]_{2\pi}\right]_{2\pi}.\qquad
\end{IEEEeqnarray}
We note that if
\begin{IEEEeqnarray}{rCl}\label{eq:conditionDelta}
\left|\frac{2\pi f_{\mathrm{c}}^\mathrm{D}}{c}(r_{m,\ell}(nT)-r_{m,\ell}((n-1)T))\right|<\pi
\end{IEEEeqnarray}
then
\begin{IEEEeqnarray}{rCl}\label{eq:downlink_delta}
\Delta^\mathrm{D}_{m,\ell}[n]= -\frac{2\pi f_{\mathrm{c}}^\mathrm{D}}{c}(r_{m,\ell}(nT)-r_{m,\ell}((n-1)T))
\end{IEEEeqnarray}
and hence equation \eqref{eq:explicit_UL} can be written as
\begin{IEEEeqnarray}{rCl}\label{eq:UL_from_DL}
\measuredangle h^\mathrm{U}_{m,\ell}[n]&=&\left[\Theta^\mathrm{U}_{m,\ell}[0]+\frac{f_{\mathrm{c}}^\mathrm{U}}{f_{\mathrm{c}}^\mathrm{D}}\sum_{k=1}^n\Delta^\mathrm{D}_{m,\ell}[k]\right]_{2\pi}.
\end{IEEEeqnarray}
Thus, Equation \eqref{eq:UL_from_DL} establishes the pseudo channel reciprocity by showing that (given \eqref{eq:conditionDelta}) the uplink can be estimated from the downlink.

\begin{figure}[t]
\begin{center}
\includegraphics[width=110mm]{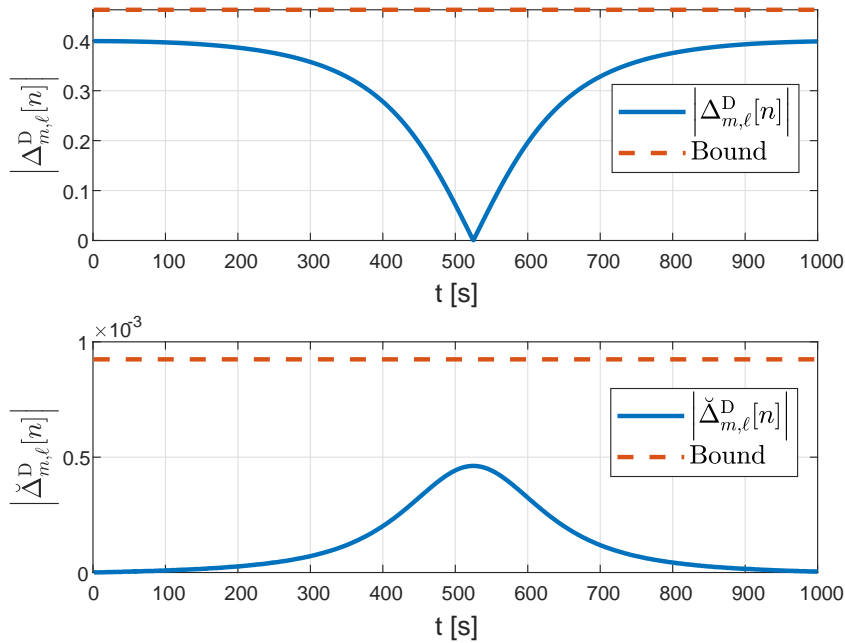}
    \caption{Phase difference for consecutive samples, $\Delta^\mathrm{D}_{m,\ell}[n]$, and phase difference for consecutive samples relative to the reference antenna,  $\breve \Delta^\mathrm{D}_{m,\ell}[n]$, for the entire  LEO satellite motion, for a satellite at a height of  $1000$km, with a transmit rate of $10$MHz and a carrier frequency of $30$GHz. Top plot shows the naive reciprocity scheme described in Section \ref{sec:pseudo} (see \eqref{eq:UL_from_DL}). Bottom plot shows the practical reciprocity scheme described in Section \ref{sec:freq_offset_not_zero} (see \eqref{eq:breve_UL_from_DL}).}
\label{fig:delta_bound}
\end{center}
\end{figure}

To show that  condition \eqref{eq:conditionDelta} is reasonable in practical cases,  consider a satellite with orbit radius $R$ and  orbit period $T_\mathrm{o}$. By bounding the phase increments using   the maximal Doppler shift at each instant, \( n\), one obtains
\begin{IEEEeqnarray}{rCl}\label{eq:delta_upper_bound}
\hspace{-5mm}\left|\Delta^\mathrm{D}_{m,\ell}[n]\right|&=&\left|\frac{2\pi f^\mathrm{D}_\mathrm{c}}{c} [r_{m,\ell}((n-1)T)-r_{m,\ell}(nT)]\right|=\left|\frac{2\pi f^\mathrm{D}_\mathrm{c}v_{m,\ell}[n]T}{c} \right|\le\left|\frac{2\pi f^\mathrm{D}_\mathrm{c}2\pi T R}{cT_\mathrm{o}} \right|\qquad
\end{IEEEeqnarray}
where $v_{m,\ell}[n]={r_{m,\ell}((n-1)T)-r_{m,\ell}(nT)}/{T}$ is the   velocity at time $n$ of satellite $\ell$ relative to antenna $m$ at the mobile.  The inequality states that the relative velocity is bounded by the satellite's orbital velocity,  $v_{m,\ell}[n]\leq\frac{2\pi R}{T_{\mathrm{o}}}$.  For illustration, consider a LEO satellite  traveling  $1000$km above the LT while transmitting at a bandwidth of $10$MHz with a $30$GHz carrier frequency. In this case we have $T=10^{-7}_s, T_0=6.298\cdot10^{3}_s$ and
\begin{IEEEeqnarray}{rCl}\label{eq:upper_bound_delta}
\left|\Delta^\mathrm{D}_{m,\ell}[n]\right|\leq\left|\frac{2\pi f^\mathrm{D}_\mathrm{c}2\pi T R}{cT_\mathrm{o}} \right|=0.46.
\end{IEEEeqnarray}
The top part of Fig. \ref{fig:delta_bound} depicts $\Delta^\mathrm{D}_{m,\ell}[n]$ from  satellite rise to satellite set, from the LT  perspective. The figure shows that  \eqref{eq:upper_bound_delta} is satisfied for every $n$, and is tightest near the rise and set of the satellite.
In the remainder of this paper we assume that  \eqref{eq:conditionDelta} holds.

\subsection{Estimation Error}
We now also consider  the presence of estimation errors, and examine the performance of the proposed scheme.

\begin{definition}\label{def:def1}
Given the initial estimate \( \hat \Theta^\mathrm{U}_{m,\ell}[0]\)  and the  sequence of estimates \( \measuredangle \hat  h^\mathrm{D}_{m,\ell}[i]\) for \( i=0,...,n\), we define the recursive estimation procedure as
\begin{IEEEeqnarray}{rCl}
\label{eq:tracking_formula}
\measuredangle \hat  h^\mathrm{U}_{m,\ell}[n]&=&\left[\measuredangle \hat h^\mathrm{U}_{m,\ell}[n-1]+\frac{f_{\mathrm{c}}^\mathrm{U}}{f_{\mathrm{c}}^\mathrm{D}}\hat \Delta^\mathrm{D}_{m,\ell}[n]\right]_{2\pi}
\end{IEEEeqnarray}
where \(\hat \Delta^{\rm D}_{m,\ell}[n] =\left[\measuredangle\hat h_{m,\ell} ^{\rm D}[n]- \measuredangle\hat h_{m,\ell} ^{\rm D}[n-1]\right]_{2\pi} \).
\end{definition}
As  explained above, the amplitude of the channel changes very slowly; hence we assume that the LT has exact knowledge of $\mathbf{D}^\mathrm{D}[n]$. Thus, we only consider the error in the estimation of $\mathbf{H}^\mathrm{D}[n]$. Let,
\begin{IEEEeqnarray}{rCl}\label{eq:dpll_input_h}
\hat h^\mathrm{D}_{m,\ell}[n]=h^\mathrm{D}_{m,\ell}[n]+w^\mathrm{D}_{m,\ell}[n]
\end{IEEEeqnarray}
where $w^\mathrm{D}_{m,\ell}[n]\sim \mathcal{CN}(0,\sigma^2)$ is the estimation error. Thus, the phase increment is given by
\begin{IEEEeqnarray}{rCl}\label{eq:Delta_m_l_n}
\hat\Delta^\mathrm{D}_{m,\ell}[n]&=&-\left[\left[2\pi f_{\mathrm{c}}^\mathrm{D}r_{m,\ell}(nT)/c+\zeta^\mathrm{D}_{m,\ell}[n]\right]_{2\pi}-\left[2\pi f_{\mathrm{c}}^\mathrm{D}r_{m,\ell}((n-1)T)/c+\zeta^\mathrm{D}_{m,\ell}[n-1]\right]_{2\pi}\right]_{2\pi}\qquad
\end{IEEEeqnarray}
where $\zeta^\mathrm{D}_{m,\ell}[n]=\measuredangle \hat  h^\mathrm{D}_{m,\ell}[n]-\measuredangle   h^\mathrm{D}_{m,\ell}[n]$ is the downlink-phase estimation error.
Here, we extend \eqref{eq:conditionDelta} to assume that
\begin{IEEEeqnarray}{rCl}
\label{eq:reciprocity_condition}
\left|\frac{2\pi f^\mathrm{D}_\mathrm{c}}{c} [-r_{m,\ell}((n)T)+r_{m,\ell}((n-1)T)]-\zeta^\mathrm{D}_{m,\ell}[n]+\zeta^\mathrm{D}_{m,\ell}[n-1]\right|<\pi.
\end{IEEEeqnarray}
Thus, \eqref{eq:UL_from_DL} becomes
\begin{IEEEeqnarray}{rCl}
\label{eq:h_n_uplink_up}
\measuredangle \hat h^\mathrm{U}_{m,\ell}[n]
&=&\left[\measuredangle h^\mathrm{U}_{m,\ell}[n]+\zeta^\mathrm{U}_{m,\ell}[0]+\frac{f_{\mathrm{c}}^\mathrm{U}}{f_{\mathrm{c}}^\mathrm{D}}\left(\zeta^\mathrm{D}_{m,\ell}[n]-\zeta^\mathrm{D}_{m,\ell}[0]\right)\right]_{2\pi}\label{eq:estimation_D_with_error}
\end{IEEEeqnarray}
where $\zeta^\mathrm{U}_{m,\ell}[0]=\measuredangle \hat  h^\mathrm{U}_{m,\ell}[0]-\measuredangle   h^\mathrm{U}_{m,\ell}[0]$
is the  uplink-phase  estimation error at $n=0$. Thus, the error in $\measuredangle \hat h^\mathrm{U}_{m,\ell}[n]$  is a linear combination of  $\zeta^\mathrm{U}_{m,\ell}[0],\zeta^\mathrm{D}_{m,\ell}[0]$ and $,\zeta^\mathrm{D}_{m,\ell}[n]$; i.e., for every $n$, the cumulative error is composed of only three elements. This is a desired estimation, as we can track the uplink for very long time, without accumulation of noise.

\par

Thus, the main problem in the implementation of this scheme is the probability that  \eqref{eq:reciprocity_condition} will not hold. In this case,  $\measuredangle \hat h^\mathrm{U}_{m,\ell}[n]$ has an additional error of $\pm 2\pi k \frac{f_{\mathrm{c}}^\mathrm{U}}{f_{\mathrm{c}}^\mathrm{D}}$, for some $ k\in \mathbb{N}$.  This estimation error is typically large, and cannot be recovered later, except by additional feedback. Thus, it is important to study and characterize these ``jump'' events. In the following subsection we present an improved tracking approach using \gls{dpll}, and study the probability of "jump" events (which are termed {\em cycle slip} in the context  of DPLL).

\subsection{DPLL Pseudo Reciprocity and Cycle-Slip}
The well-known \gls{dpll} is very similar in structure to the scheme in definition \ref{def:def1}, except for the multiplication of the channel phase increments by a different constant. These different multiplicative constants aim to achieve different  goals. In traditional DPLL this multiplication enables further noise reduction by taking advantage of the correlation between successive increments. In our approach, the multiplication by a constant achieves the pseudo reciprocity by converting the downlink phase increments into uplink phase increments.
\par
Many algorithms and works show good performance using DPLL for phase tracking (see for example \cite{golshan2007pulse,van2013single,fitz1995performance,zhuang1996performance}). Thus, we present an extended DPLL scheme that aims to achieve both the noise reduction and the pseudo reciprocity.


\tikzstyle{block} = [draw, fill=white, rectangle,
minimum height=3em, minimum width=6em]
\tikzstyle{smallblock} = [draw, fill=white, rectangle,
minimum height=2em, minimum width=3em]
\tikzstyle{sum} = [draw, fill=white, circle, node distance=1cm]
\tikzstyle{input} = [coordinate]
\tikzstyle{output} = [coordinate]
\tikzstyle{pinstyle} = [pin edge={to-,thin,black}]

\begin{figure}[t]
\begin{center}
\begin{tikzpicture}[auto, node distance=2cm,>=latex']

\node [input, name=input] {};
\node [sum, right of=input, node distance=1.5cm] (sum) {};
\draw (sum.north east) -- (sum.south west) (sum.north west) -- (sum.south east);
\node [block, right of=sum, node distance=3.5cm] (Ablock) {Sinusoidal Phase Detector};
\draw [->] (sum) -- node[name=a] {} (Ablock);

\node [output, above of = a, node distance=1.7cm, xshift=11.1cm] (output1) {};
\node [output, below of = output1, node distance=4.85cm, xshift=0cm] (output2) {};
\node [block, below of = a, node distance=2.15cm, xshift=6.1cm] (Cblock) {VCO};
\node [smallblock, above of = a, node distance=1.7cm, xshift=6.1cm] (Bblock) {$\frac{f_\mathrm{c}^\mathrm{U}}{f_\mathrm{c}^\mathrm{D}}$};
\node [block, above of = a, node distance=1.7cm, xshift=8.6cm] (Dblock) {VCO};
\node [smallblock, below of=sum, node distance=3cm] (Eblock) {$e^{-j(\cdot)}$};
\draw [draw,->] (input) -- node {$\hat h^\mathrm{D}_{m,\ell}[n]$} (sum);

\draw [->] (Eblock) -- node[name=Es] {} (sum);

\draw [->] (Ablock) -| node {$\tilde\phi[n]$}(Cblock){};
\draw [->] (Bblock) -- node {}(Dblock){};
\draw [->] (Dblock) -- node [pos=1.6]{$\hat\Theta^\mathrm{U}_{m,\ell}[n]\qquad\qquad\qquad$}(output1){};
\draw [->] (Cblock) |- node [pos=1.07]{$\hat\Theta^\mathrm{D}_{m,\ell}[n]\qquad\qquad\qquad$}(output2){};
\draw [->] (Cblock) |- node {}(Eblock){};
\draw [->] (Ablock) -| (Bblock);

\end{tikzpicture}
\caption{Block diagram of the proposed DPLL pseudo reciprocity}
\label{fig:pll}
\end{center}
\end{figure}
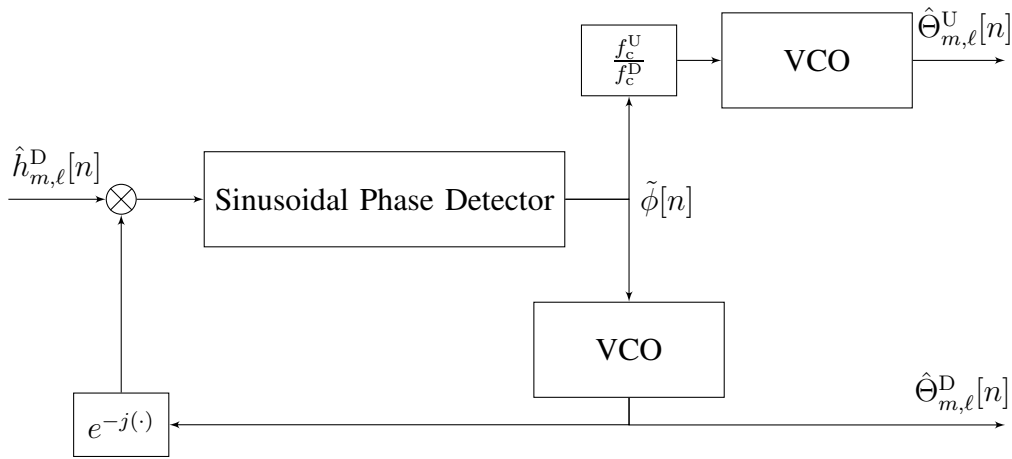

 We use a first-order DPLL that consists of two voltage-controlled-oscillators (VCO), one for the downlink phase estimation and one for the uplink, as depicted in Fig. \ref{fig:pll}. The  output of the downlink  VCO  is used to close the DPLL loop. It also produces the improved channel phase estimate, $\hat\Theta^\mathrm{D}_{m,\ell}[n]$. The second VCO is used to generate the uplink estimates, using the pseudo reciprocity increments.
\par
The VCO has a delay of one sample time; i.e., $\hat\Theta^\mathrm{D}_{m,\ell}[n]$ is the estimation of the input at instant $n$ based on samples up to $n-1$. The DPLL input is given in \eqref{eq:dpll_input_h}, and the {\em sinusoidal phase extractor}  approximates the input phase by its imaginary part, given by
\begin{IEEEeqnarray}{rCl}
\tilde\phi[n]&=&\Im(h^\mathrm{D}_{m,\ell}[n]e^{-j\hat\Theta^\mathrm{D}_{m,\ell}[n]}+e^{-j\hat\Theta^\mathrm{D}_{m,\ell}[n]} w^\mathrm{D}_{m,\ell}[n])\notag\\
&=&|h^\mathrm{D}_{m,\ell}[n]|\sin(\Theta^\mathrm{D}_{m,\ell}[n]-\hat\Theta^\mathrm{D}_{m,\ell}[n])+\tilde w^\mathrm{D}_{m,\ell}[n]\label{eq:tile_phi}
\end{IEEEeqnarray}
where $\tilde w^\mathrm{D}_{m,\ell}[n]=\Im(e^{-j\hat\Theta^\mathrm{D}_{m,\ell}[n]}w^\mathrm{D}_{m,\ell}[n])\sim\mathcal{N}(0,\sigma^2/2)$ and $\Im(\cdot)$ represents the imaginary part.\footnote{The phase  can also be taken as the angle of the complex number. However, this phase extraction changes the error distribution,  which makes the cycle slip analysis much more complicated. Note that numerical simulations show that in the low SNR regime, a complex phase extractor does not improve the DPLL tracking performance compared to the sinusoidal phase extractor.}

\par
The downlink channel phase estimates are the output of the first VCO
\begin{IEEEeqnarray}{rCl}
\hat\Theta^\mathrm{D}_{m,\ell}[n]&=&KT\tilde\phi[n-1]+\hat\Theta^\mathrm{D}_{m,\ell}[n-1]\label{eq:Phi^O_k}\\
&=&\sum_{k=1}^{n-1}KT\tilde\phi[k]+\hat\Theta^\mathrm{D}_{m,\ell}[0]\notag
\end{IEEEeqnarray}
whereas the  uplink channel phase estimates are the output of the second VCO\begin{IEEEeqnarray}{rCl}\label{dpll:uplink_est}
\hat\Theta^\mathrm{U}_{m,\ell}[n]&=&\frac{f_\mathrm{c}^\mathrm{U}}{f_\mathrm{c}^\mathrm{D}}\sum_{k=1}^{n-1}KT\tilde\phi[k]+\hat\Theta^\mathrm{U}_{m,\ell}[0].
\end{IEEEeqnarray}

\begin{figure}[t]
\begin{center}
\includegraphics[width=110mm]{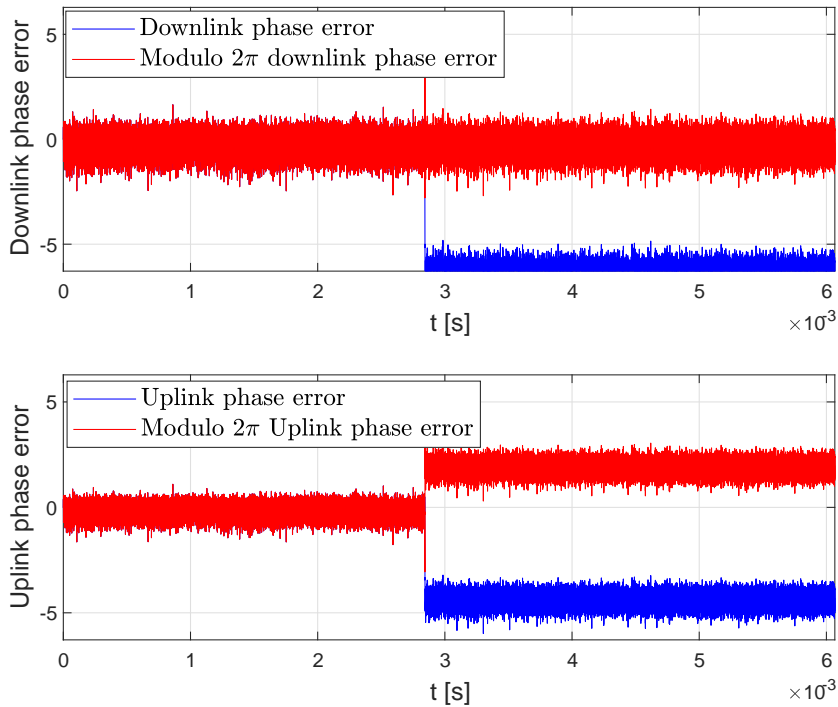}
    \caption{Channel estimation using the proposed scheme $f_\mathrm{c}^\mathrm{U}=20 $MHz, $f_\mathrm{c}^\mathrm{U}=30$MHz. The figure shows the estimation in the downlink (top) and uplink (bottom). In each channel, we plot the error (in blue) and its modulo $2\pi$ (in red).}
\label{fig:DL_UL_errors}
\end{center}
\end{figure}

Using our single channel feedback (for each antenna) we set $\hat\Theta^\mathrm{U}_{m,\ell}[0]=\Theta^\mathrm{U}_{m,\ell}[0]$. Thus, we can denote the  downlink channel phase estimation error of the DPLL at sample $n$ as
\begin{IEEEeqnarray}{rCl}\label{breve_phi}
\phi[n]&=&\Theta^\mathrm{D}_{m,\ell}[n]-\hat \Theta^\mathrm{D}_{m,\ell}[n].
\end{IEEEeqnarray}
Note that both $\Theta^\mathrm{D}_{m,\ell}[n]$ and $\hat\Theta^\mathrm{D}_{m,\ell}[n]$  are not wrapped by modulo (i.e., they can take values outside the range $[-\pi,\pi]$).
\begin{definition}
Given the error $\phi[n]$ at each sample $n$, we define a
cycle-slip as the event where
\begin{IEEEeqnarray}{rCl}\label{eq:cyc_cond}
\left|\phi[n]\right|\geq\pi.
\end{IEEEeqnarray}
\end{definition}
Note that \eqref{eq:cyc_cond} is the complementary condition to \eqref{eq:reciprocity_condition}. Thus, the pseudo reciprocity holds as long as there is no cycle slip. To illustrate the pseudo reciprocity sensitivity to a cycle-slip, we consider an antenna of a  single satellite  and a single antenna at the  LT. The LT uses DPLL to estimate the downlink and uplink channel phases based on the downlink channel pilot sequences. Fig. \ref{fig:DL_UL_errors} shows the estimation errors for the downlink and for the uplink. In each direction, we plot two errors, the actual error ($\phi[n]$ in blue) and the modulo $2\pi$ error (in red). As shown by the actual downlink error (in blue), a cycle-slip occurs in the downlink channel phase, around time $t=2.8\cdot10^{-3}_s$. As a result, the actual error of the uplink also suffers from a cycle-slip event. The difference between the downlink and uplink channel phases performance is seen in the modulo $2\pi$ error. The cycle-slip event does not affect the estimation of the downlink channel phases, but does affect the estimation of the uplink channel phases by a significant factor of $-2\pi\frac{f_\mathrm{c}^\mathrm{U}}{f_\mathrm{c}^\mathrm{D}}\approx -4.18$.

\par
To sum up, if the LT successively keeps track of the phase variations between consecutive times, then losing a $2\pi$-cycle only occurs in the case of a cycle slip. As long as there is no cycle sleep, the uplink channel estimation is proportional to the downlink channel error (which is typically very small). In what follows, we derive the \gls{mnsll} in DPLL, which we harness to show that the reciprocity can hold for very long times.

\subsection{Cycle-Slip Analysis}\label{sec:Performance_analysis}
We now derive an approximation to the \gls{mnsll} for the proposed system (cf. Fig \ref{fig:pll}). We start by rewriting  $\phi[n]$, by substituting  \eqref{eq:tile_phi} and \eqref{eq:Phi^O_k} into \eqref{breve_phi}:
\begin{IEEEeqnarray}{rCl}\label{xxx}
\phi[n]&=&\Theta^\mathrm{D}_{m,\ell}[n]-\hat\Theta^\mathrm{D}_{m,\ell}[n-1]-KT|h^\mathrm{D}_{m,\ell}[n]|\sin(\phi[n-1])\\
&&-KT\tilde w^\mathrm{D}_{m,\ell}[n-1]-\Theta^\mathrm{D}_{m,\ell}[n-1]+\Theta ^\mathrm{D}_{m,\ell}[n-1]\notag
\end{IEEEeqnarray}
where we also add and subtract $\Theta^\mathrm{D}_{m,\ell}[n-1]$. This analysis is done under the assumption of sufficiently high sampling frequency. Thus, it is reasonable to neglect
\begin{IEEEeqnarray}{rCl}\label{eq:zero_doppler}
\Theta^\mathrm{D}_{m,\ell}[n]-\Theta^\mathrm{D}_{m,\ell}[n-1]=0.
\end{IEEEeqnarray}
Using \eqref{breve_phi} for $\phi[n-1]$ we can write
\begin{IEEEeqnarray}{rCl}\label{eq:phi_digital}
\frac{\phi[n]-\phi[n-1]}{T}&=&-KA\sin(\phi[n-1])-K\tilde w^\mathrm{D}_{m,\ell}[n-1]
\end{IEEEeqnarray}
where $A=|h^\mathrm{D}_{m,\ell}[n]|.$
\par
Viterbi \cite{viterbi1963phase} calculated the mean time for cycle slip in PLL where the input is a continuous real signal. To  use the Viterbi results, note that the stochastic difference equation \eqref{eq:phi_digital} can be seen as the Euler approximation \cite{kloeden1992applications} of the stochastic differential equation
\begin{IEEEeqnarray}{rCl}\label{eq:phi_analog}
{\rm d }\phi (t)&=&-KA\sin(\phi(t)) {\rm d}t-K\tilde w^\mathrm{D}_{m,\ell}(t){\rm d}t.
\end{IEEEeqnarray}
Furthermore,  the diffusion function,  $-K$, and the drift function, $-AK\sin(\cdot)$, are infinitely  continuously differentiable. Hence the process $\phi[n]$ in \eqref{eq:phi_digital} converges weakly to  $\phi(t)$ in \eqref{eq:phi_analog}.

Viterbi, calculated the mean time to lose lock for equation \eqref{eq:phi_analog}, which is given by
\begin{IEEEeqnarray}{rCl}\label{eq:viterbi_res}
T_{\mathrm{cys}}&=&\frac{8\pi^2}{\sigma^2K^2}J_0^2(j\alpha)
\end{IEEEeqnarray}
where
\begin{IEEEeqnarray}{rCl}\label{eq:effctive_snr}
\alpha=\frac{4A}{K\sigma^2}
\end{IEEEeqnarray}
is the effective SNR and $J_0(\cdot)$ is the Bessel function of the first kind. By comparing \eqref{eq:phi_digital} to \eqref{eq:phi_analog}, Equation \eqref{eq:viterbi_res} can we converted to its discrete equivalent. Thus, the MSL can be approximated by
\begin{IEEEeqnarray}{rCl}\label{eq:msl}
N_{\mathrm{cys}}&\approx&\frac{8\pi^2}{\sigma^2K^2T}J_0^2(j\alpha).
\end{IEEEeqnarray}
\par
Naturally, the MSL decreases monotonically with $K$ and $T$. However, the choice of $K$ and $T$ must also take into account the tracking ability and the robustness to noise. The effect of these two quantities can be quantified for example through the \gls{mse} of the steady state. As we will show, the MSL is typically very large, thus in the numerical section we chose $K$ to minimize the steady state MSE. See  appendix \ref{app:K} for details.

The MSL is also affected by the effective SNR, $\alpha$. Note that $J_0(j\alpha)$ is a monotonically increasing function in $\alpha$. Furthermore, for large $\alpha$, we can approximate $J_0(j\alpha)\sim\exp(\alpha)/\sqrt{2\pi\alpha}$; thus, a slight improvement in the SNR can cause a significant improvement in the MSL.

\section{Reciprocity with Frequency offset}\label{sec:freq_offset_not_zero}
The previous section assumed no frequency offset between the satellites and the LT ($f_\ell^\mathrm{D}=0$). We next present an improved reciprocity procedure that works with  $f_\ell^\mathrm{D}\neq0$. To do so, we rewrite the channel matrices, which involves normalizing the channel gains by the phase of antenna $1$. That is,  we rewrite the  downlink channel matrix as $\mathbf{\tilde H}^\mathrm{D}[n]=\mathbf{\breve H}^\mathrm{D}[n]\mathbf{\breve D}^\mathrm{D}[n]$, where
\begin{IEEEeqnarray}{rCl}
\hspace{-1mm}\mathbf{\breve D}^\mathrm{D}[n]=\begin{bmatrix}\frac{\Upsilon^\mathrm{D}_{1}}{r_{1,1}(nT)}e^{j2\pi f_1^\mathrm{D}nT}e^{j\alpha^\mathrm{D}_1}e^{j2\pi f\mathrm{_c^D}  \frac{r_{1,1}(nT)}{c}}   & \cdots & 0 \\
\vdots\ & \ddots & \vdots \\
0 & \cdots & \frac{\Upsilon^\mathrm{D}_{L}}{r_{1,L}(nT)}e^{j2\pi f_L^\mathrm{D}nT}e^{j\alpha^\mathrm{D}_L}  e^{j2\pi f\mathrm{_c^D}  \frac{r_{1,L}(nT)}{c}} \\
\end{bmatrix}\notag
\end{IEEEeqnarray}

\begin{IEEEeqnarray}{rCl}
\mathbf{\breve H}^\mathrm{D}[n]=\begin{bmatrix}1 & \cdots & 1\\
e^{j\breve\Theta^\mathrm{D}_{2,1}[n]}  & \cdots & e^{j\breve\Theta^\mathrm{D}_{2,L}[n]} \\
\vdots\ & \ddots & \vdots \\
e^{j\breve\Theta^\mathrm{D}_{M,1}[n]} & \cdots & e^{j\breve\Theta^\mathrm{D}_{M,L}[n]}  \\
\end{bmatrix}
\end{IEEEeqnarray}
and
\begin{IEEEeqnarray}{rCl}\label{eq:breve_dl_phi}
\breve\Theta^\mathrm{D}_{m,\ell}[n]=-\frac{2\pi (f^{\mathrm{D}}_\ell+f_{\mathrm{c}}^\mathrm{D})(r_{m,\ell}(nT)-r_{1,\ell}(nT))}{c}.
\end{IEEEeqnarray}
In the same manner, we rewrite the  uplink channel matrix as $\mathbf{\tilde H}^\mathrm{U}[n]=\mathbf{H}^\mathrm{U}[n]\mathbf{D}^\mathrm{U}[n]$, where
\begin{IEEEeqnarray}{rCl}
\hspace{-1mm}\mathbf{\breve D}^\mathrm{U}[n]=\begin{bmatrix}\frac{\Upsilon^\mathrm{U}_{1}}{r_{1,1}(t)}e^{-j2\pi f_1^\mathrm{U}nT}e^{j\alpha^\mathrm{U}_1}e^{j2\pi f\mathrm{_c^U}  \frac{r_{1,1}(nT)}{c}}  & \cdots & 0 \\
\vdots\ & \ddots & \vdots \\
0 & \cdots & \frac{\Upsilon^\mathrm{U}_{L}}{r_{1,L}(t)}e^{-j2\pi f_L^\mathrm{U}nT}e^{j\alpha^\mathrm{U}_L}e^{j2\pi f\mathrm{_c^U}  \frac{r_{1,L}(nT)}{c}}  \\
\end{bmatrix}\notag
\end{IEEEeqnarray}

\begin{IEEEeqnarray}{rCl}
\hspace{-1mm}\mathbf{\breve H}^\mathrm{U}[n]=\begin{bmatrix} 1 & e^{j\breve\Theta^\mathrm{U}_{1,2}[n]} &\cdots & e^{j\breve\Theta^\mathrm{U}_{1,L}[n]}\\
\vdots &\vdots & \ddots & \vdots \\
1  & e^{j\breve\Theta^\mathrm{U}_{M,2}[n]} &\cdots & e^{j\breve\Theta^\mathrm{U}_{M,L}[n]}\\
\end{bmatrix}\notag
\end{IEEEeqnarray}
and
\begin{IEEEeqnarray}{rCl}\label{eq:breve_ul_phi}
\breve\Theta^\mathrm{U}_{m,\ell}[n]=-\frac{2\pi f_{\mathrm{c}}^\mathrm{U}(r_{m,\ell}(nT)-r_{1,\ell}(nT))}{c}.
\end{IEEEeqnarray}
Matrices $\mathbf{\breve D}^\mathrm{U}[n]$ and $\mathbf{\breve D}^\mathrm{D}[n]$ are diagonal. Thus, both the ZF precoder and the ZF equalizer only require  knowledge of the matrices $\mathbf{\breve H}^\mathrm{U}[n]$ and $\mathbf{\breve H}^\mathrm{D}[n]$, respectively.
\par
Note that all the elements in the matrices $\mathbf{\breve H}^\mathrm{U}[n]$ and $\mathbf{\breve H}^\mathrm{D}[n]$ have an amplitude of $1$. Thus,  it is sufficient to study their phases.  Moreover, inspection of \eqref{eq:breve_dl_phi} and \eqref{eq:breve_ul_phi} reveals that they do not depend on the frequency error. Thus, this reciprocity scheme works as well for $f^\mathrm{D}_\ell\ne 0$. In addition, \eqref{eq:breve_dl_phi} and \eqref{eq:breve_ul_phi} are only affected  form the differential Doppler between the two antennas (as seen from the subtraction $r_{m,\ell}(nT) -r_{1,\ell}(nT)$). Since the antennas are relatively close, the differential Doppler is significantly smaller than the actual Doppler shift. Hence, the phase tracking in this scheme is much simpler.
\par

Nevertheless, this scheme has two disadvantages, both related to using  measurements that are phase differences between two antennas. The first disadvantage is that the noise is stronger than the phase estimation over a single antenna. However, below we show that the noise increase is negligible compared to the benefit of much slower tracking. The second disadvantage is an
error distribution that is more complicated than in the previous section. This will prevent us from    presenting an exact analytical study. Instead, we will present a good approximation  based on a Gaussian error distribution.
\par
To better explain this improved reciprocity scheme, we define the range differences in time and space:
\begin{IEEEeqnarray}{rCl}\label{eq:breve_r}
\breve d_{m,\ell}[n]&=&r_{m,\ell}((n-1)T)-r_{1,\ell}((n-1)T)-r_{m,\ell}(nT)+r_{1,\ell}(nT)
\end{IEEEeqnarray}
and
\begin{IEEEeqnarray}{rCl}
\measuredangle\breve h^\mathrm{U}_{m,\ell}[n]&=&\left[\breve\Theta^\mathrm{U}_{m,\ell}[n]\right]_{2\pi}=\left[\breve\Theta^\mathrm{U}_{m,\ell}[0]+\frac{2\pi f_{\mathrm{c}}^\mathrm{U}}{c}\sum_{k=1}^n\breve d_{m,\ell}[k]\right]_{2\pi}\label{eq:explicit_UL_part2}.
\end{IEEEeqnarray}
Thus, to apply reciprocity, we need to estimate $2\pi f_c^\mathrm{U} \breve d_{m,\ell}[n]/c $ from the downlink.
\par
The  downlink channel phase in this model is
\begin{IEEEeqnarray}{rCl}\label{xxx}
\measuredangle\breve  h^\mathrm{D}_{m,\ell}[n]&=&\left[\breve\Theta^\mathrm{D}_{m,\ell}[n]\right]_{2\pi}=\left[-2\pi (f_\ell^\mathrm{D}+f_{\mathrm{c}}^\mathrm{D})(r_{m,\ell}(nT)-r_{1,\ell}(nT))/c\right]_{2\pi}
\end{IEEEeqnarray}
and the downlink measured phase change is $\breve \Delta^\mathrm{D}_{m,\ell}[n]=\left[\measuredangle \breve h^\mathrm{D}_{m,\ell}[n]-\measuredangle \breve h^\mathrm{D}_{m,\ell}[n-1]\right]_{2\pi}$. Thus, in the absence of noise, if
\begin{IEEEeqnarray}{rCl}\label{eq:condition_reciprocity2}
\left|\frac{2\pi (f_\ell^\mathrm{D}+f_{\mathrm{c}}^\mathrm{D})}{c}\breve d_{m,\ell}[n]\right|<\pi
\end{IEEEeqnarray}
we can write
\begin{IEEEeqnarray}{rCl}\label{eq:breve_delta}
\breve \Delta^\mathrm{D}_{m,\ell}[n]&=&\frac{2\pi (f_\ell^\mathrm{D}+f_{\mathrm{c}}^\mathrm{D})}{c}\breve d_{m,\ell}[n]\approx\frac{2\pi f_{\mathrm{c}}^\mathrm{D}}{c}\breve d_{m,\ell}[n]
\end{IEEEeqnarray}
where the approximation comes from the reasonable assumption that $f_\ell^\mathrm{D}\ll f_{\mathrm{c}}^\mathrm{D}$. Hence, equation \eqref{eq:explicit_UL_part2} can be written as
\begin{IEEEeqnarray}{rCl}\label{eq:breve_UL_from_DL}
\measuredangle \breve h^\mathrm{U}_{m,\ell}[n]&=&\left[\breve\Theta^\mathrm{U}_{m,\ell}[0]+\frac{f_{\mathrm{c}}^\mathrm{U}}{f_{\mathrm{c}}^\mathrm{D}}\sum_{k=1}^n\breve \Delta^\mathrm{D}_{m,\ell}[k]\right]_{2\pi}
\end{IEEEeqnarray}
which shows that the reciprocity works even with \( f_{\ell}^{\rm D}\neq 0\) and with a minimal feedback of $\breve\Theta^\mathrm{U}_{m,\ell}[0]$. Note that the condition in \eqref{eq:condition_reciprocity2} is over changes in the distance difference. As $\left|r_{m,\ell}(nT)-r_{m,\ell}((n-1)T)\right|\gg\left|\breve d_{m,\ell}[n]\right|$, this constraint is much more relaxed than the constraint in \eqref{eq:conditionDelta}. In appendix \ref{doppler_part2_bound} we show that \eqref{eq:condition_reciprocity2} holds for any practical system. To make this advantage clearer, the top of Fig. \ref{fig:delta_bound} depicts $\breve\Delta^\mathrm{D}_{m,\ell}[n]$ for an example system setup (the top of the figure depicts  $\Delta^\mathrm{D}_{m,\ell}[n]$ for the same setup). It is obvious that $\breve\Delta^\mathrm{D}_{m,\ell}[n]$ is significantly smaller than $\Delta^\mathrm{D}_{m,\ell}[n]$. These slow variations in the phase difference (i.e., small $\breve\Delta^\mathrm{D}_{m,\ell}[n]$) can be used to improves the tracking, complexity and the \gls{mnsll} of the system.

The cycle slip analysis in the previous section does not apply to the scheme of this section because the estimation noise is no longer Gaussian (because  we multiply each element $\hat h^\mathrm{D}_{m,\ell}[n]$ by the conjugate of the first antenna $\hat h^\mathrm{D}_{1,\ell}[n]$). Nevertheless, we found that the cycle slip analysis can produce an excellent approximation for this scheme by simply adjusting  the  noise variance. As in the previous section, the conditional distribution of each channel estimation is assumed to be $\hat h^\mathrm{D}_{m,\ell}[n]\Big|h^\mathrm{D}_{m,\ell}[n]\sim \mathcal{CN}(h^\mathrm{D}_{m,\ell}[n],\sigma^2)$ (see Eq. \eqref{eq:dpll_input_h}). Thus, for the improved scheme, the variance at the input to the DPLL  is given by
\begin{IEEEeqnarray}{rCl}\label{xxx}
\breve\sigma^2&=&\mathrm{Var}\left[\hat h^{*\mathrm{D}}_{1,\ell}[n]\hat h^\mathrm{D}_{m,\ell}[n]\Big|h^\mathrm{D}_{1,\ell}[n],h^\mathrm{D}_{m,\ell}[n]\right]\notag\\
&=&\mathbb{E}\Bigg[|h^\mathrm{D}_{1,\ell}[n]|^2|h^\mathrm{D}_{m,\ell}[n]|^2+|w^\mathrm{D}_{1,\ell}[n]|^2|w^\mathrm{D}_{m,\ell}[n]|^2+|h^\mathrm{D}_{m,\ell}[n]|^2|w^\mathrm{D}_{1,\ell}[n]|^2\notag\\
&&+|h^\mathrm{D}_{1,\ell}[n]|^2|w^\mathrm{D}_{m,\ell}[n]|^2\Bigg|h^\mathrm{D}_{1,\ell}[n],h^\mathrm{D}_{m,\ell}[n]\Bigg]-\left|\mathbb{E}\left[h^{*\mathrm{D}}_{1,\ell}[n]h^{\mathrm{D}}_{m,\ell}[n]\Big|h^\mathrm{D}_{1,\ell}[n],h^\mathrm{D}_{m,\ell}[n]\right]\right|^2\notag\\
&=&\sigma^4+2\sigma^2\label{eq:new_var}
\end{IEEEeqnarray}
where we used $|h^\mathrm{D}_{m,\ell}[n]|^2=1$ (see Eq. \eqref{eq:H_D}).

Substituting \eqref{eq:new_var} into \eqref{eq:msl}, the MSL is approximated by
\begin{IEEEeqnarray}{rCl}\label{eq:msl2}
N_{\mathrm{cys}}&\approx&\frac{8\pi^2}{(\sigma^4+2\sigma^2)K^2T}J_0^2\left(j\frac{4A}{K(\sigma^4+2\sigma^2)}\right).
\end{IEEEeqnarray}
In Fig. \ref{fig:cyc_sim_vs_viterbi_part2} we show that \eqref{eq:msl2} is an excellent approximation for the MSL of the improved scheme.

\begin{figure}[t]
\begin{center}
\includegraphics[width=110mm]{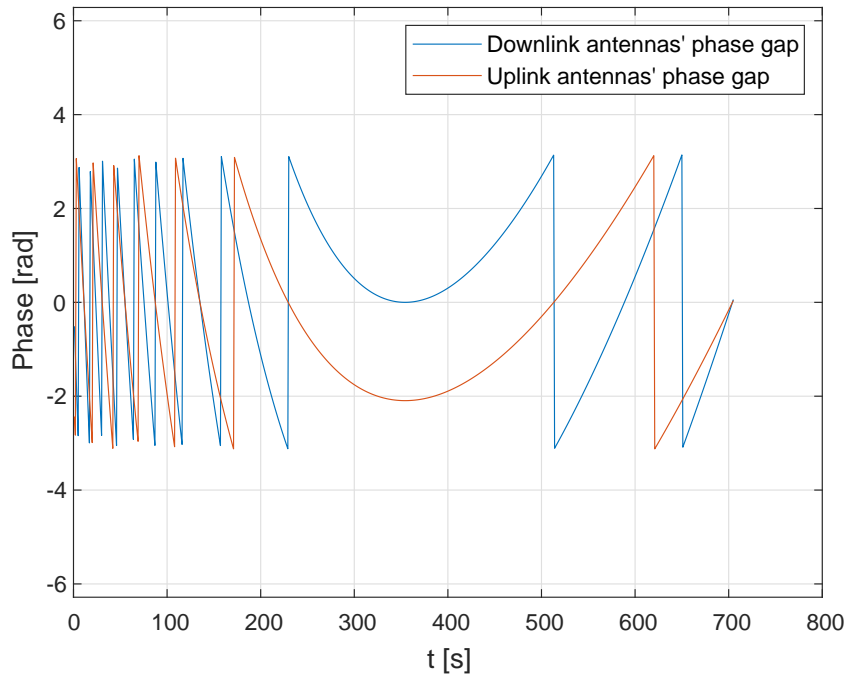}
    \caption{The phase gap between two LT antennas in the  downlink channel phase and uplink channel phase over time where one antenna satellite communicates with two  antennas LT.}
\label{fig:phase_diff}
\end{center}
\end{figure}

\section{Numerical Results}
The previous sections described our reciprocity procedure in FDD satellite communication system and derived an approximation to the \gls{mnsll}, which indicates the reciprocity failure probability. Here we present numerical results showing these expressions' usefulness and accuracy.
\par
In Fig. \ref{fig:phase_diff}, we consider a single-antenna satellite at a  height of  $1000$km  that communicates with a double antenna LT ($d=50$cm separation between the antennas). The carrier frequencies are $30$GHz and $20$GHz for the downlink and uplink channels, respectively. To illustrate the challenge of exploiting this uplink-downlink reciprocity, we plot the downlink and the uplink phase gaps between the LTs antennas. We aim to show no remarkable information between the downlink and uplink phase gaps. Because one may obtain that our reciprocity procedure is not innovative and can be understood quite straightforwardly. Fig. \ref{fig:phase_diff} shows that this interpretation is incorrect; the correlation between the downlink and uplink phase gaps is very pure and changes over time.\par
\begin{figure}[t]
\begin{center}
\includegraphics[width=110mm]{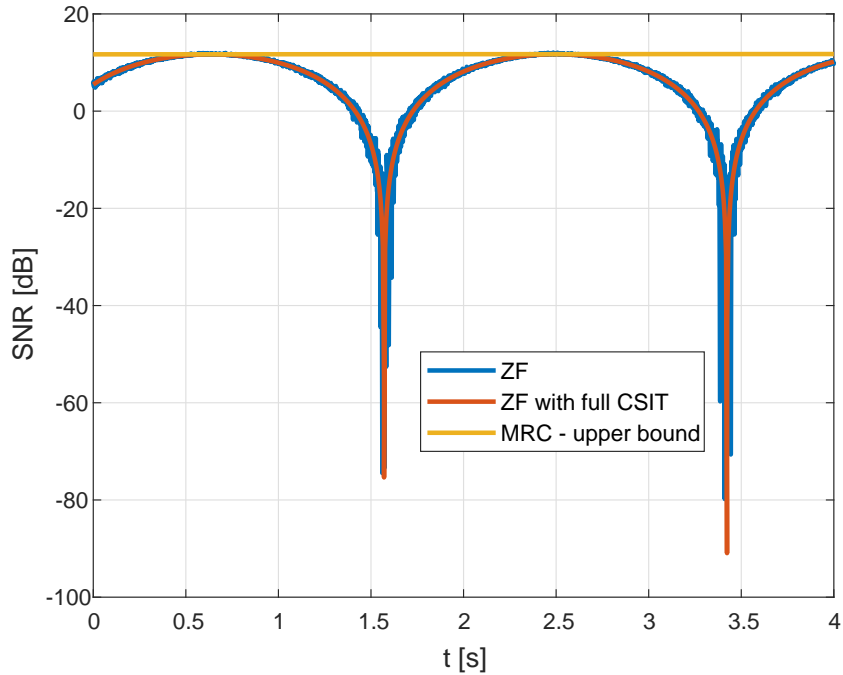}
    \caption{LT with two transmit antennas send independent data streams simultaneously to two satellites with one receive antenna. The figure plots the receive signal SNR at one satellite for three precoding cases: the first is ZF-precoding with full CSI, the second is ZF-precoding without CSI and with reciprocity channel estimation. The third is for comparison, where the LT sends only data to one satellite having MRC precoding with half of the power compared to the first two cases.}
\label{fig:snr_vs_t_part2}
\end{center}
\end{figure}

 Fig. \ref{fig:snr_vs_t_part2} considers an FDD  satellite communication system with two single-antenna satellites  communicating with a double-antenna LT. The satellites send downlink pilots periodically to the LT, whereas in the uplink, the LT sends independent data streams simultaneously to each satellite. The LT tracks the downlink channel phase variations over time and estimates the uplink channel phase to send the two data streams. We used  typical   SatCom parameters. The downlink and uplink carrier frequencies are $30$GHz and  $20$GHz, respectively. In both channels we used: a s  $45$dBi transmit antenna gain, a   $20$dBi receive antenna gain, a $42$dBm transmit power, a $10$MHz bandwidth, and a noise power spectral density of $\sigma^2=-170$dBm/Hz.

\par
Fig. \ref{fig:snr_vs_t_part2} depicts the receive  SNR at one satellite for three  cases. In the first one, the LT has perfect  CSIT whereas the second case involves an imperfect CSIT where the LT achieves CSIT using the proposed reciprocity scheme. In both cases, the LT sends an independent data stream to each satellite simultaneously using ZF-precoding. For comparison, we consider a third case where the LT sends data to only one of the satellites using MRC precoding. Here, the LT uses its two antennas to transmit with half of the power compared to the first two cases. The figure shows that the LT estimates  the uplink channel phase successfully using reciprocity with approximately the same performance as in the case of full CSIT. However, the singular values of the uplink channel phase matrix vary with time. Occasionally the matrix is well-conditioned, and the ZF-precoding works well, whereas at other times the matrix is ill conditioned  so that canceling the interference to each satellite is impossible, even in the full CSIT case.
\par
\begin{figure}[t]
\begin{center}
\includegraphics[width=110mm]{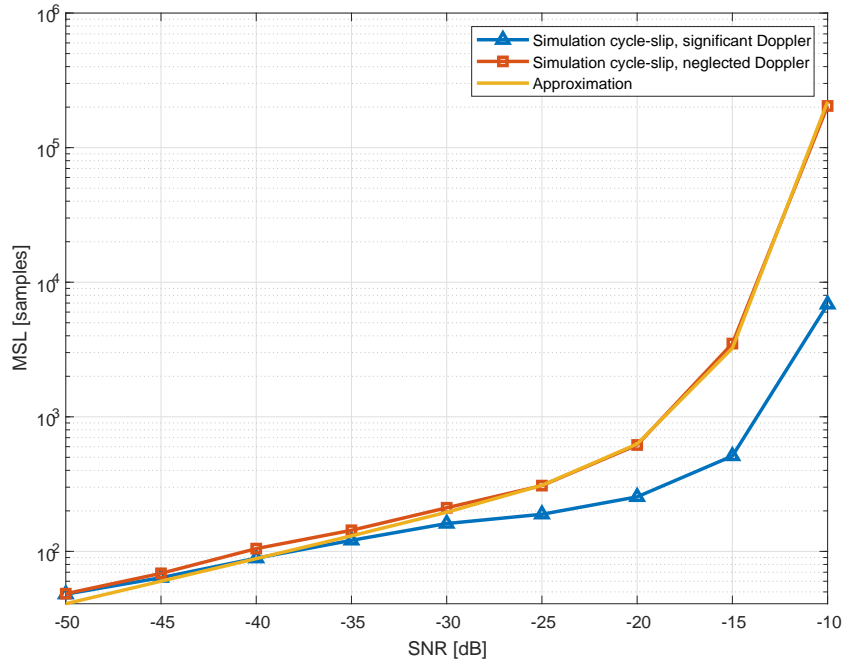}
    \caption{The simulation of the first cycle-slip in the downlink channel and the theoretical cycle-slip approximation  as a function of the received signal SNR where one antenna satellite transmits to one receive antenna LT.}
\label{fig:cyc_sim_vs_viterbi}
\end{center}
\end{figure}
\begin{figure}[t]
\begin{center}
\includegraphics[width=110mm]{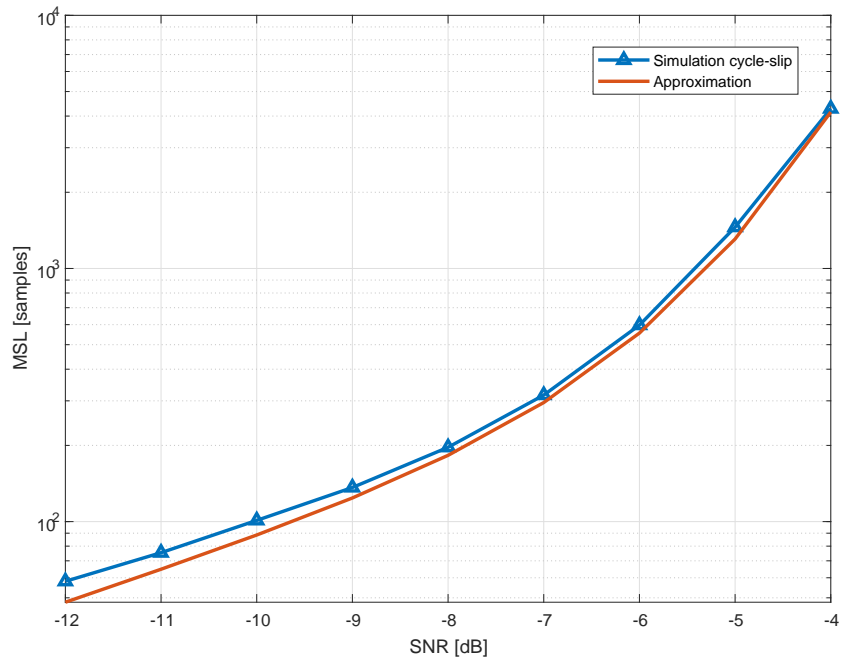}
    \caption{The simulation \gls{mnsll} in the downlink channel and the theoretical \gls{mnsll} as a function of the received signal SNR where one satellite  transmits to LT with two receive antennas}
\label{fig:cyc_sim_vs_viterbi_part2}
\end{center}
\end{figure}

To show a match to the theory, Figs. \ref{fig:cyc_sim_vs_viterbi} and \ref{fig:cyc_sim_vs_viterbi_part2} simulate the received signal at a very low SNR, even though this SNR range is not typical of satellite communication systems. The reason is that in the standard receive SNR range, the \gls{mnsll} is enormous and difficult to simulate. It is important to emphasize that our \gls{mnsll} formula \eqref{eq:msl} holds for all SNR regimes. We denote the received SNR for both cases as $1/\sigma^2$.

In Fig. \ref{fig:cyc_sim_vs_viterbi}, we consider one single-antenna satellite transmitting to a single antenna LT with a $100$MHz bandwidth, a $100$MHz sample rate ($T=1\cdot10^{-8}_s$), and a carrier frequency of $30$GHz. The LT tracks  the downlink channel phase  and estimates the uplink channel phase using reciprocity (cf. section \ref{sec:pseudo}). Fig. \ref{fig:cyc_sim_vs_viterbi} depicts the mean of the first  cycle-slip in the downlink channel and the theoretical cycle-slip approximation \eqref{eq:msl} as a function of the received  SNR. The figure depicts two cases: the first with a significant Doppler shift and the second with a neglected Doppler effect; each case corresponds to a  specific LT-satellite instantaneous configuration. In both cases, in low SNR, the DPLL struggles to track the channel phase, so that a cycle slip occurs early. As the SNR grows, the DPLL tracking  improves and the \gls{mnsll} increases. Because the \gls{mnsll} formula \eqref{eq:msl}  considers a zero Doppler shift, the presence of such may cause a cumulative phase error, which explains the early cycle slip in the high Doppler case.\par
In Fig. \ref{fig:cyc_sim_vs_viterbi_part2}, we consider the same setup as in Fig. \ref{fig:cyc_sim_vs_viterbi}, but with two antennas at the LT. Here, the LT  tracks  the downlink channel phase variations between the receive antennas and consecutive times (see \eqref{eq:breve_delta} and \eqref{eq:breve_r}). The satellite transmission parameters consist of a: bandwidth of $10$Hz , a sample rate of $10$Hz ($T=0.1_s$) and carrier frequency of $30$GHz. The downlink channel phase-variations between receive antennas contain a very small Doppler; hence, there is no cumulative Doppler phase error at the DPLL. Fig. \ref{fig:cyc_sim_vs_viterbi_part2} depicts the actual downlink \gls{mnsll}, obtained via Monte-Carlo, and the approximate \eqref{eq:msl2}  as a function of the received SNR. There is a good match between the simulation and the theory. The small Doppler shift enables the LT to reduce the DPLL gain $K$ hence improving the effective SNR  of the DPLL. Moreover, this  small  Doppler enables the LT to reduce the sampling rate (only $10$Hz!!) and still successfully track the uplink channel. According the theoretical and the numerical results, in typical SatCom SNRs, the LT usually does not experience cycle slips for the entire satellite LOS motion.

\section{Conclusions}
In this paper we showed the pseudo reciprocity of the uplink and downlink in FDD LOS channels.  We focused on  LEO SatCom systems, which are typically based on FDD and exhibit LOS dominant channels. We presented a novel scheme which allows  the LT to track phase changes in the downlink channel and use them to estimate the phase changes in the uplink channel, effectively estimating the CSIT by tracking the CSIR. Our proposed pseudo-reciprocity approach achieves CSIT estimation with minimal feedback and system complexity.

Incorporating our innovative approach into SatCom systems has the potential to substantially enhance the implementation of MU-MIMO technology in SatCom. This improvement, in turn, could lead to a significant increase in data rates. The key to this enhancement lies in the fact that pseudo-reciprocity facilitates the estimation of CSIT for a large number of channels, thereby streamlining the process and improving overall performance.

We also analyzed the impact of estimation errors in the CSIR on the estimation errors in the CSIT. Our results indicate that there is no cumulative error over time, ensuring that pseudo-reciprocity remains stable for extended periods. Effectively, in most scenarios it will be sufficient to send feedback once when the satellite rises and then track the channel with small errors until the satellite sets.
\par 
To track changes in the CSIR, we proposed using a DPLL. The  DPLL introduces a cycle slip probability, prompting us to calculate the Mean Time of the First Cycle Slip (MSL).

In the numerical results section, we demonstrated that using pseudo-reciprocity for precoding achieves performance comparable to precoding with full CSIT knowledge. Additionally, we confirmed that the MSL calculation aligns with our simulations. We found that within the common Signal-to-Noise Ratio (SNR) range in SatCom, cycle slips are so rare  that they are not likely to occur for any satellite motion, from the satellite rise to set. Thus, our proposed pseudo-reciprocity approach offers a promising avenue for enhancing satellite communication systems while maintaining minimal complexity and feedback requirements.

\begin{appendices}

\section{Continuous process and digital noise comparison}\label{app:noise}
In this Appendix we prove that the limit of $\tilde w^\mathrm{D}_{m,\ell}[n]$ as $T$ goes to zero is a white Gaussian noise with PSD of $\sigma^2/2$.

Let $g(t)$ be a general continuous function. We denote $\tilde w^\mathrm{D}_{m,\ell}(t)$ as a white Gaussian process with zero mean and PSD $\sigma^2/2$. In order to sample the white process, we denote the filter
\begin{IEEEeqnarray}{rCl}\label{xxx}
f(t)&=W\cdot1_{[-1/2W,1/2W]}(t)
\end{IEEEeqnarray}
and
\begin{IEEEeqnarray}{rCl}\label{xxx}
\tilde w^\mathrm{D}_{m,\ell}[n]&=&\tilde w^\mathrm{D}_{m,\ell}(t)*f(t)\Big|_{t=nM/W}
\end{IEEEeqnarray}
where $W$ is the bandwidth and $T=M/W$ is the sampling gap, $M\in\mathbb{N}$.

We want to show that $\tilde w^\mathrm{D}_{m,\ell}[n]$ has the same distribution as $\tilde w^\mathrm{D}_{m,\ell}(t)*f(t)$. The distribution of $\tilde w^\mathrm{D}_{m,\ell}[n]$ is a white Gaussian with zero mean and variance $\mathrm{Var}[\tilde w^\mathrm{D}_{m,\ell}[n]]=\sigma^2W/2$.
Next, we calculate
\begin{IEEEeqnarray}{rCl}\label{xxx}
\mathrm{Var}\left[\int g(t)\tilde w^\mathrm{D}_{m,\ell}(t)dt\right]&=&\sigma^2/2\int g^2(t)dt
\end{IEEEeqnarray}
 and show  that where $T$ goes to zero
\begin{IEEEeqnarray}{rCl}\label{xxx}
\lim_{W\rightarrow\infty,n/W\rightarrow t}\mathrm{Var}\left[\sum_n g(nT)\tilde w^\mathrm{D}_{m,\ell}[n]/W\right]&=&\sigma^2/2\int g^2(t)dt
\end{IEEEeqnarray}
which ends the proof.

\section{Calculate $K$ that minimizes the MSE}\label{app:K}
We show one way to choose $K$ that minimizes the specific downlink \gls{mse}, which is given in \eqref{breve_phi}. We assume that the changes between two consecutive distinct downlink times is $f_dT$ where $f_d$ is the maximum Doppler and $T$ is the sampling gap. We also use the small angle approximation $\sin(\phi[n-1])\approx\phi[n-1]$. Based on these assumptions,
\begin{IEEEeqnarray}{rCl}\label{xxx}
\phi[n]&=&f_dT-KTA\phi[n-1]-KT\tilde w^\mathrm{D}_{m,\ell}[n-1]+\phi[n-1]\notag
\end{IEEEeqnarray}
and the MSE is given by
\begin{IEEEeqnarray}{rCl}
\mathrm{MSE}[\phi[n]]&=&\mathrm{Var}[\phi[n]]+\mathbb{E}^2[\phi[n]]\\
&=&(1-KTA)^2\mathrm{Var}[\phi[n-1]]+\mathrm{Var}[KT\tilde w^\mathrm{D}_{m,\ell}[n-1]]+((1-KTA)\mathbb{E}[\phi[n-1]]+f_dT)^2\notag.
\end{IEEEeqnarray}

Assuming stability, $\mathrm{Var}[\phi[n]]=\mathrm{Var}[\phi[n-1]]$ and $\mathbb{E}[\phi[n]=\mathbb{E}[\phi[n-1]$, therefore
\begin{IEEEeqnarray}{rCl}
\mathrm{MSE}[\phi[n]]&=&\frac{\sigma^2KAT}{A^2(2-KAT)}+\frac{T^2f_d^2}{(KAT)^2}\\
\end{IEEEeqnarray}
and the $K$ value that gives the minimum is the solution of
\begin{IEEEeqnarray}{rCl}
0&=&(KAT)^3\gamma-T^2f_d^2(KAT)^2+4T^2f_d^2(KAT)-4T^2f_d^2.
\end{IEEEeqnarray}

\section{Doppler upper bound}\label{doppler_part2_bound}
\begin{figure}[t]
\begin{center}
\includegraphics[width=100mm]{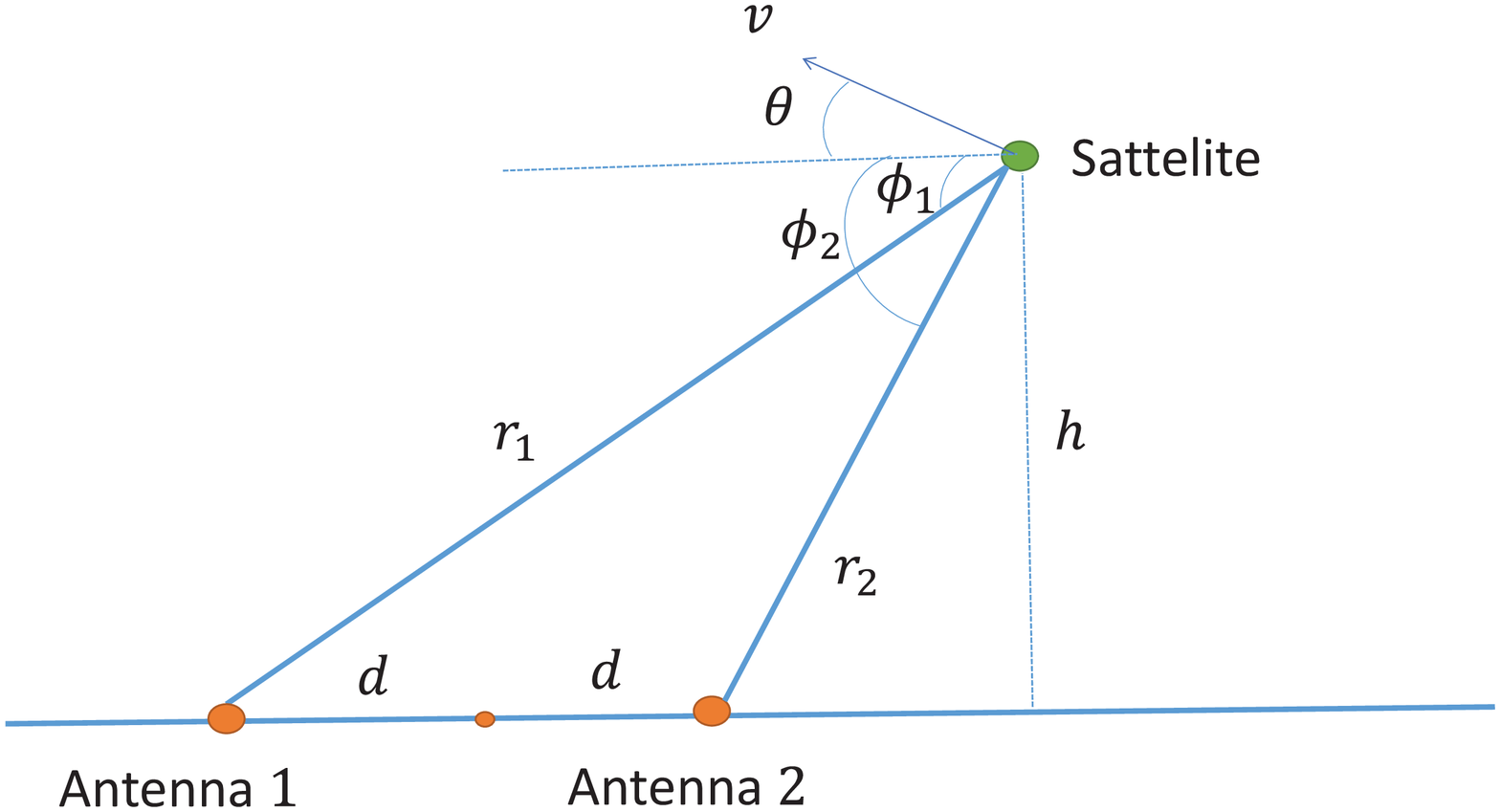}
    \caption{The general setup of two receive antennas on Earth and one antenna satellite. We analyze this setup to prove that \eqref{eq:condition_reciprocity2} is satisfied.}
\label{fig:doppler_bound}
\end{center}
\end{figure}

In this Appendix we calculate an upper bound for the absolute value of the Doppler shift between the consecutive times and between antennas to prove that \eqref{eq:condition_reciprocity2} is satisfied. We consider the general setup of two receive antennas on Earth and one antenna satellite as depicted in Fig. \ref{fig:doppler_bound}. We denote $v$ as the satellite velocity, $h$ as the satellite height, $r_1$ and $r_2$ are the distances between the satellite and antennas $1$ and $2$, respectively, $d$ is the distance between the antennas, and $\phi_1$,$\phi_2$ and $\theta$ as defined in Fig. \ref{fig:doppler_bound}.

The Doppler shift absolute value variation over time between antenna $1$ and antenna $2$ is given by

\begin{IEEEeqnarray}{rCl}\label{xxx}
\left|\Delta f\right|&=&\left|\frac{vf_c}{c}\cos(\theta)[\cos(\phi_1)-\cos(\phi_2)]\right|\notag\\
&=&\left|\frac{vf_c}{c}\cos(\theta)[-2\sin(\frac{\phi_1+\phi_2}{2})\sin(\frac{\phi_1-\phi_2}{2})]\right|\notag\\
&\leq&\left|\frac{2vf_c}{c}\sin(\frac{\phi_1-\phi_2}{2})\right|.
\end{IEEEeqnarray}

Using the sinusoidal formula on the triangle we can find that $\sin(\phi_2-\phi_1)=\frac{d\sin(\phi_1)}{r_1}$ and  assuming $|\phi_1-\phi_2|<\pi/2$, we can write

\begin{IEEEeqnarray}{rCl}\label{xxx}
\left|\Delta f\right|&\leq&\left|\frac{2vf_c}{c}\sin(\phi_1-\phi_2)\right|\notag\\
&=&\left|\frac{2dvf_c}{cr_1}\sin(\phi_1)\right|\notag\\
&\leq&\frac{2dvf_c}{ch}\label{eq:appx_bound}
\end{IEEEeqnarray}

Using \eqref{eq:appx_bound}, we can write
\begin{IEEEeqnarray}{rCl}
\left|\frac{2\pi f_{\mathrm{c}}^\mathrm{U} \breve d_{m,\ell}(n)}{c}\right|=\left|\frac{2\pi f_{\mathrm{c}}^\mathrm{U} \breve d_{m,\ell}(n)T}{cT}\right|\leq\left|\frac{4\pi d f_{\mathrm{c}}^\mathrm{U} vT}{ch}\right|
\end{IEEEeqnarray}

and for all common satellites orbits, carrier frequencies and sampling rates $\left|\frac{4\pi d f_{\mathrm{c}}^\mathrm{U} vT}{ch}\right|<\pi$.

\end{appendices}
\bibliography{ReiBib}
\bibliographystyle{ieeetr}
\end{document}